\documentclass[12pt]{article}
\usepackage[margin=1in]{geometry}
\usepackage{setspace}
\usepackage{amsmath}
\usepackage{enumerate}
\usepackage{epsfig}
\usepackage{color}
\usepackage{amssymb}
\usepackage{multirow}
\usepackage{epstopdf}
\newtheorem{thm}{Theorem}[section]

\numberwithin{equation}{section}

\title{A two-component normal mixture alternative to the Fay-Herriot model}

\begin{document}
\maketitle
\centerline{\bf
 Adrijo Chakraborty
 \footnote{NORC at the University of Chicago, Bethesda, MD 20814, chakraborty-adrijo@norc.org},
 Gauri Sankar Datta
 \footnote{Department of Statistics,
            University of Georgia, Athens, GA 30602, USA. email:gauri@stat.uga.edu}
 \footnote{Center for Statistical Research and Methodology, US Census Bureau, Washington, D.C. 20233}
 \footnote{Disclaimer: This report is released to inform interested parties of research and to encourage discussion of work in progress. The views expressed are those of the authors and not necessarily those of the US Census Bureau}
            and Abhyuday Mandal
 \footnote{Department of Statistics,
            University of Georgia, Athens, GA 30602, USA. email:amandal@stat.uga.edu}}

\vskip .5in
\noindent
{\bf Abstract:} This article considers a robust hierarchical Bayesian approach to deal with random effects of small area means when some of these effects assume extreme values, resulting in outliers. In presence of outliers, the standard Fay-Herriot model, used for modeling area-level data, under normality assumptions of the random effects may overestimate random effects variance, thus provides less than ideal shrinkage towards the synthetic regression predictions and inhibits borrowing information. Even a small number of substantive outliers of random effects result in a large estimate of the random effects variance in the Fay-Herriot model, thereby achieving little shrinkage to the synthetic part of the model or little reduction in posterior variance associated with the regular Bayes estimator for any of the small areas. While a scale mixture of normal distributions with known mixing distribution for the random effects has been found to be effective in presence of outliers, the solution depends on the mixing distribution. As a possible alternative solution to the problem, a two-component normal mixture model has been proposed based on noninformative priors on the model variance parameters, regression coefficients and the mixing probability. Data analysis and simulation studies based on real, simulated and synthetic data show advantage of the proposed method over the standard Bayesian Fay-Herriot solution derived under normality of random effects.

\section{Introduction}

\noindent Small area estimation methods are getting increasingly popular among survey practitioners. Reliable small area estimates are often solicited by the policy makers from both government and private sectors for planning, marketing and decision making. In order to support growing demand of reliable small area estimates, researchers have developed methods that combine information from the small areas and other related variables. Ghosh and Rao (1994), Rao (2003), Jiang and Lahiri (2006), Datta (2009) and Pfeffermann (2013) provided a comprehensive review of the research in small area estimation.

\bigskip \noindent The landmark paper by Fay and Herriot (1979) used the empirical Bayes (EB) approach (see, for example, Efron and Morris, 1973) and popularized model-based small area estimation methods. Denoting the design-based direct survey estimator of the $i$th small area by $Y_i$ and its auxiliary variable by $x_i$, an $r\times 1$ vector, Fay and Herriot (1979) introduced the model
\begin{eqnarray}\label{model:eq1}
Y_i = \theta_i + e_i, \quad \theta_i = x^T_i\beta+v_i, \quad  i = 1,\ldots, m.
\end{eqnarray}
Here $\theta_i$ is a summary measure of the characteristic to be estimated for the $i$th small area, $e_i$ is the sampling error of the  estimator $Y_i$, and the random effects $v_i$ denotes the model error measuring the departure of $\theta_i$ from its linear regression on $x_i$. It is assumed that $e_1,\ldots,e_m$ are independent and normally distributed with $e_i\sim N(0,D_i)$, and are independent of $v_1,\ldots,v_m$, which are i.i.d. $N(0,A)$. The sampling variances $D_i$'s are treated as known, but the model parameters $\beta$ and $A$ are unknown. Random effects $v_i$'s are also known as small area effects.

\bigskip \noindent In this paper we focus on hierarchical Bayes (HB) methods for area-level models. The classical area-level Fay-Herriot model was primarily developed as a frequentist model, which was later given a Bayesian formulation (Rao 2003; Datta et al. 2005). Estimators obtained from Fay-Herriot model are shrinkage estimators, i.e., an weighted average of the direct estimator and the model-based synthetic estimator, these weights depend on the model assumption. Datta and Ghosh (2012) gave an extensive review of shrinkage estimation in small area estimation context. Shrinkage estimators are primarily constructed to improve the standard estimators. {For instance, in small area context model based shrinkage estimators are constructed to improve the precision of direct estimators such as {the sample mean or Horwitz-Thompson estimator}.} Datta and Lahiri (1995) discussed how outliers can affect shrinkage estimators, even a single outlier may lead all the small area estimates to collapse to their corresponding direct estimates. This phenomenon was also mentioned in the context of estimation of multiple normal means under the assumption of an exchangeable normal prior (cf. Efron and Morris 1971, Stein 1981, and Angers and Berger 1991). One or more substantive outliers considerably inflate a standard estimator of model variance.

\bigskip \noindent Overestimation of model variance due to one or more substantive outliers practically results in no shrinkage of any of the direct estimates of the small area means to a synthetic regression estimator. This would also limit the reduction of the posterior variances of the model-based estimates. To rectify this problem, following the work of Angers and Berger (1991), who used a Cauchy distribution for the small area means $\theta_i$, Datta and Lahiri (1995) recommended a broader class of heavy-tailed distributions through scale mixture of normal distributions. They showed that under these assumptions, in presence of substantive outliers, estimators corresponding to the outlying areas converge to their corresponding direct estimators but leave the non-outlying areas less affected. One difficulty with the last method is that the mixing distribution for the scale parameter is considered to be known.  For example, one can use $t$-distribution for random effects as in Xie et al.  (2007). However, in absence of any information regarding the degrees of freedom one needs to specify a prior. Xie et al. (2007) assumed a gamma prior for the degrees of freedom. The hyperparameters involved in this gamma distribution need to be specified. Bell and Huang (2006) argued that under practical circumstances limited information is obtained from the data regarding degrees of freedom and instead they used several fixed values for the degrees of freedom.

\bigskip \noindent In order to avoid specifying the mixing distribution {mentioned in the last paragraph}, in this paper we propose a two-component normal mixture distribution for the random small area effects. {Our model accommodates means for} outlying areas to come from a distribution with the {larger} variance. It is a simple extension of the Fay-Herriot model with a contaminated random effects distribution with possibly a small proportion of areas having a larger model variance. Contaminated models have been extensively used in empirical evaluations of robust empirical best linear unbiased prediction (EBLUP) approach of Sinha and Rao (2009).
 We consider an HB approach by assigning non-subjective priors to the parameters involved in the model. Some components of these priors are improper, hence we provide sufficient conditions for the posterior distribution to be proper.

\bigskip \noindent In a recent article, Datta et al. (2011) demonstrated that in the presence of good covariates $x_i$, the variability of the small area means $\theta_i$ may be accounted well by $x_i$, and including a random effects $v_i$ in the model (\ref{model:eq1}) may be unnecessary. These authors test a null hypothesis of no random effects in the small area model and if it is not rejected, they propose more accurate synthetic estimators for the small area means. In a more recent article, Datta and Mandal (2015) argued that even if the null hypothesis is rejected in this case, it is reasonable to expect only a small fraction of small areas means will not be adequately explained by covariates, and only these areas would require a random component to the regression model.

\bigskip \noindent Using the HB approach, Datta and Mandal (2015) considered a ``spike and slab'' distribution for the random small area effects in order to propose a flexible balance between Fay and Herriot (1979) and Datta et al. (2011) models. However,  often it is difficult to find reliable covariates that would describe the response well, particularly, if the number of small areas is large. For such datasets, not only the test proposed by Datta et al. (2011) would suggest an inclusion of small area effects, the model proposed by Datta and Mandal (2015) would also estimate the probability of existence of random effects to be as very high. This effectively would suggest the Fay-Herriot model, but in reality, only a small proportion of small areas may not be adequately explained by a model with one single $A$. This would result in  overestimation of $A$, thereby resulting in a poor fit, particularly when the number of small areas $m$ is large. Even if most of the small areas would require a random effects term in the regression model, it is more likely that only a small proportion of small areas would need a bigger value of $A$, and a smaller value of the same is sufficient for the other areas. In this paper, we assume that $v_1,\ldots, v_m$ are independently distributed with mean 0 and a two-component mixture of normal distributions with variance either $A_1$ or $A_2 (>A_1)$. This model is potentially useful to handle large outliers in small area means.

\bigskip\noindent Bell and Huang (2006) presented an insightful discussion about using $t$-distribution with known d.f. to handle outliers in Fay-Herriot model. The theoretical regression residuals from (\ref{model:eq1}) consist of the sum of the sampling error and model error, which are not individually observable. They argued that a residual may be an outlier either due to the sampling error or the model error. It is difficult to distinguish between the scenarios of a sampling error outlier or a model error outlier since the data in fitting the model  (\ref{model:eq1}) cannot readily disentangle the two cases. They explained that consequences of these two types of outliers are quite different. If the model error $v_i$ is an outlier for some area, then the regression model (or synthetic estimation) is not good for the area. In that case, the direct estimator $Y_i$ should be used as the small area estimator. Datta and Lahiri (1995) considered this case using a scale mixture of normal distribution. An alternative to this approach is proposed in the present article through a two-component normal mixture. Bell and Huang (2006) noted that in the presence of a model outlier, if the direct estimator also has large variability, then no satisfactory solution exists.
On the other hand, if the sampling error $e_i$ is an outlier due to an underestimation of the variance $D_i$, then the direct estimator $Y_i$ is not reliable; Bell and Huang (2006) argued that the ``synthetic estimator'' $x_i^T\beta$ may be used for prediction. To address this issue, they proposed a $t$-distribution for the sampling distribution.  For further discussion, we refer to this article.

\bigskip\noindent {There is a substantive literature on frequentist approach for robust estimation of small area means in presence of outliers. Ghosh et al. (2008) considered robust empirical Bayes estimation of small area means for area level model. They used Huber's $\psi$-function to limit influence of outliers. For unit level model Sinha and Rao (2009) and Chambers et al. (2014) proposed robust modification of EBLUPs of finite population means of small areas. They also used Huber's $\psi$-function to limit the impact of outlier observations on the estimators of model parameters and the best linear unbiased predictors. While Sinha and Rao (2009) provided  robust projective EBLUPs (in the terminology of Chambers et al. (2014)) of finite population small area means, the latter group of authors discussed the limitation of such predictors in terms of bias, and also proposed robust predictive EBLUPs to remedy this concern. }

\bigskip \noindent This paper is organized as follows.  In Section~\ref{sec:ch2model} we describe the proposed model and discuss some properties of our new shrinkage estimators. In Section~\ref{sec:ch2da} we illustrate our method to estimate U.S. poverty rates for 3141 counties based on 5-year estimates from the American Community Survey.  Performance of the model in comparison with the traditional Fay-Herriot model is discussed in Section~\ref{sec:ch2sim} and Section~\ref{sec:performance}. Section~ \ref{Discussion} provides a concluding discussion. A detailed proof of the propriety of the posterior distribution is relegated to the Appendix.

\section{Two-component normal mixture model}\label{sec:ch2model}

Fay and Herriot (1979) proposed a model which has been extensively used in many small area estimation applications to provide reliable estimates of poverty and income measures. While for regular data the model successfully produces accurate shrinkage estimators of small area means, it breaks down in the presence of substantial outliers among the small area means. In order to account for the outliers, we consider a two-component normal mixture extension of Fay-Herriot model. This model is given by
\begin{align}
\label{eqn:ch2model}
y_{i}&=\theta_{i}+e_{i} ,~~
\theta_{i}=x^{T}_{i}\beta + (1-\delta_{i}) v_{1i} + \delta_{i} v_{2i},~i=1,\dots,m,
\end{align}
where $e_i$, $\delta_{i}$, $v_{1i}$, $v_{2i}$ are independently distributed with $P(\delta_{i}=1|p)=1-p$, $v_{1i}\sim N(0,A_{1})$ and $v_{2i}\sim N(0,A_{2})$. As in (\ref{model:eq1}), $\beta$ is an $r\times 1$ vector of regression parameters, and the sampling errors $e_{1},\dots,e_{m}$ are independently normally distributed.
%\bigskip \noindent
To complete our HB structure, we consider the following class of priors,
\begin{align}\label{eqn:pr1}
\pi(\beta,A_{1},A_{2},p) =\pi^*(A_1,A_2) \propto  A_{1}^{-\alpha_{1}}A_{2}^{-\alpha_{2}}I(0<A_{1}<A_{2}<\infty).
\end{align}
We use a uniform prior on the regression parameter $\beta$ and the mixing proportion $p$. For the prior on the variance parameters, we choose $\alpha_{1}<1< \alpha_{2}$ suitably, and we discuss permissible choices of the values of $\alpha_{1}$ and $\alpha_{2}$ later. We impose the restriction $A_{1}<A_{2}$, so that we do not have a label switching problem leading to non-identifiability. The area specific random effects corresponding to the outlying areas in the model are assumed to follow a normal distribution with a larger variance, which remains the motivation behind imposing such a restriction. While for the parameter $\beta$  common in all components of mixture an improper uniform prior is reasonable, the prior for $A_{1}$ and $A_{2}$, which are not common in all components of the mixing distributions, is required to be at least {\em partially proper}. By partially proper we mean that while the marginals are improper, conditional priors for $A_2$ given $A_1$, and $A_1$ given $A_2$ are proper. For this to hold for our class of priors for $A_1, A_2$, it is necessary and sufficient that $\alpha_1 < 1<\alpha_2$.  A partially proper prior is required for the parameters that are not common to all components of a Bayesian mixture model (cf. Scott and Berger, 2006).

\bigskip \noindent Since the Bayesian model involves improper priors, in Theorem \ref{thm:ch21} below we provide sufficient conditions that ensure the resulting posterior distribution from the proposed model will be proper. A detailed proof of Theorem~\ref{thm:ch21} is given in Section~\ref{sec:ch2proof}.
\begin{thm}
\label{thm:ch21}
The resulting posterior distribution from model (\ref{eqn:ch2model})  and the prior in (\ref{eqn:pr1}) will be proper if
(a) $m>r+2(2-\alpha_{1}-\alpha_{2})$ and (b)  $2-\alpha_{1}-\alpha_{2}>0$.
\end{thm}

\bigskip \noindent The sufficient conditions in Theorem~\ref{thm:ch21} provide a set of permissible values for $\alpha_{1}$ and $\alpha_{2}$. In conjunction with the condition $2-\alpha_{1}-\alpha_{2}>0$, the condition $\alpha_2 >1$ implies $\alpha_1<1$. We noted earlier that the last two conditions are necessary to elicit partially proper priors. The special case $\alpha_{1}=0$ is feasible, which corresponds to a uniform prior, provided $1<\alpha_{2}<2$. However, it is not possible to assign a uniform prior on $A_2$. If $\alpha_{1}=\frac{1}{2}$, then $1<\alpha_{2}<\frac{3}{2}$. Also, for mixture models, Jeffreys' prior has no closed-form expression to work with.

\bigskip \noindent Our choice of prior for the mixing parameter $p$ is Uniform(0,1). We can modify this prior if subjective information is available. If past experience in an application suggests any information regarding the proportion of outlying areas, that can be incorporated in the model by modifying the prior for $p$. Sufficient conditions for the propriety of the posterior density will remain unchanged. For instance, if the model is modified with the assumption that $p$ follows a known $Beta$ distribution, the sufficient conditions provided in Theorem~\ref{thm:ch21} will remain intact.

\bigskip \noindent It is well-known that even a single substantial outlier will collapse shrinkage estimators of all $\theta_i$'s based on the model (\ref{model:eq1}) to the direct estimators $y_i$'s (see Dey and Berger, 1983;  Stein, 1981). As a result, model-based estimators will fail to borrow strength from the other small areas. To protect against this odd behavior, Angers and Berger (1991) and Datta and Lahiri (1995) suggested a robust shrinkage model. These authors used suitable scale mixture of normal distributions to model long-tail distribution of the $\theta$'s. These methods assume the knowledge of the scale mixing distribution, which may not be available.
The purpose of our proposed mixture model in (\ref{eqn:ch2model}) is to provide an alternative solution that does not require the knowledge of the mixing distribution and to facilitate borrowing information among non-outlying observations in the presence of some substantive outliers.

\bigskip \noindent We discuss below a heuristic comparison of the the shrinkage property of the Bayes estimators of $\theta_i$ under the Fay-Herriot model and our proposed model, in presence of substantial outliers. For Fay-Herriot model, given the values of the parameters $\beta$ and $A$, estimator of $\theta_i$ is
\begin{eqnarray}
\theta^{FH}_{i} &=y_{i}-\dfrac{D_{i}}{D_{i}+A}(y_{i}-x^{T}_{i}\beta),~i=1,\dots, m.
\end{eqnarray}
In presence of outliers, frequentist estimators of $A$ will be large, and the posterior density of $A$ will have a long right tail, which will also result in a large Bayesian estimator of $A$. Consequently, an estimate of the shrinkage coefficient $D_i/(D_i+A)$ will be rather small, and the Bayes or the EB estimator of $\theta_i$ will borrow little from its synthetic regression prediction and it will collapse to direct estimator $y_{i}$ for all $i$.

\bigskip \noindent We now argue that the proposed mixture model is more flexible to retain shrinkage of the non-outlying observations in presence of outliers. %Again, for known model parameters $\beta, A_{1},A_{2}$ and $p$, the conditional mean
Let $E(\theta_i|\beta,A_{1},A_{2},p,y)=\theta_i^{Mix}$. % (\mbox{say})$.
Using iterated expectation  $E(\theta_i|\beta,A_{1},A_{2},p,y)= E[E(\theta_i|\beta,A_{1},A_{2},\delta_i,p,y)|\beta,A_{1},A_{2},p,y]$, and noting that $E(\theta_i|\beta,A_{1},A_{2},\delta_i,p,y)= \dfrac{D_{i}x^{T}_{i}\beta +A_{1+\delta_{i}}y_i }{D_{i}+A_{1+\delta_{i}}}$, $\tilde p_i = P(\delta_{i}=0|\beta,A_{1},A_{2},p,y)$, we get
\begin{align}
\theta^{Mix}_{i}
                 &=y_{i}-\left[ \left (\dfrac{D_{i}}{D_{i}+A_{1}} \right)\tilde p_i
                 +\left(\dfrac{D_{i}}{D_{i}+A_{2}} \right)(1-\tilde p_i)\right](y_{i}-x^{T}_{i}\beta),
\label{eq2.8}
\end{align}
where
\begin{eqnarray}
%P(\delta_{i}=1|\beta,A_{1},A_{2},p,y_{i})
\tilde p_i &=
\dfrac{\frac{p}{(D_{i}+A_{1})^{\frac{1}{2}}}\exp\left\{-\frac{1}{2}\frac{(y_{i}-x^{T}_{i}\beta)^2}{(D_{i}+A_{1})} \right\}}{\frac{p}{(D_{i}+A_{1})^{\frac{1}{2}}}\exp\left\{-\frac{1}{2}\frac{(y_{i}-x^{T}_{i}\beta)^2}{(D_{i}+A_{1})} \right\}+\frac{(1-p)}{(D_{i}+A_{2})^{\frac{1}{2}}}\exp\left\{-\frac{1}{2}\frac{(y_{i}-x^{T}_{i}\beta)^2}{(D_{i}+A_{2})} \right\}},
\end{eqnarray}
for $i=1,\dots,m$. In presence of substantially large outliers, $(y_{i}-x^{T}_{i}\beta)^{2}$ and $A_{2}$ are expected to be high, hence $P(\delta_{i}=0|\beta,A_{1},A_{2},p,y_{i}) \approx 0$. This will result in the second shrinkage term within square brackets in (\ref{eq2.8}) to be dominant. However, since the posterior distribution of $A_2$ has long tail, the shrinkage coefficient associated with the second component will be small and $\theta^{Mix}_{i} \approx y_{i}$, i.e., if the $i^{th}$ area is outlying then the small area estimator based on this model will be very close to its direct estimator. On the other hand, for any non-outlying areas $\tilde p_i$ will be away from $0$, and their shrinkages will be less impacted by the outliers.

\section{Data Analysis} \label{sec:ch2da}

\bigskip\noindent We illustrate our proposed methodology by analyzing a real data obtained from the ``American Fact Finder" website maintained by US Census Bureau. The data set contains 5-year ACS estimates of overall poverty rates for 3141 counties of United States along with their associated design-based standard errors. The county identifiers are not available due to confidentiality reasons.
In order to improve direct design-based estimates, government agencies implement state-of-the-art small area estimation methods to produce model-based estimates using auxiliary data. For poverty estimation, domain level tax data are typically used as auxiliary information. However, tax data are not available for public use owing to legal restrictions. In our analysis we use foodstamp participation rate as our only auxiliary variable (correlation between foodstamp participation rate and overall poverty rate is 0.81). Initially we fit the Fay-Herriot model (\ref{model:eq1}) with restricted maximum likelihood method (REML) as well as hierarchical Bayesian (HB) method assuming flat priors for regression and variance parameter. The REML and Bayes estimates of the model parameters are very close: $\hat{\beta}^{REML}=(0.056,0.634)^{T}$, $\hat{A}^{REML}=0.0009$ and $\hat{\beta}^{Bayes}=(0.051,0.634)^{T}$, $\hat{A}^{Bayes}=0.0009$.

\vskip .3in

\begin{figure}[h]
\begin{center}
\begin{tabular}{ccc}
        \includegraphics[width=2.5in,height=3.0in]{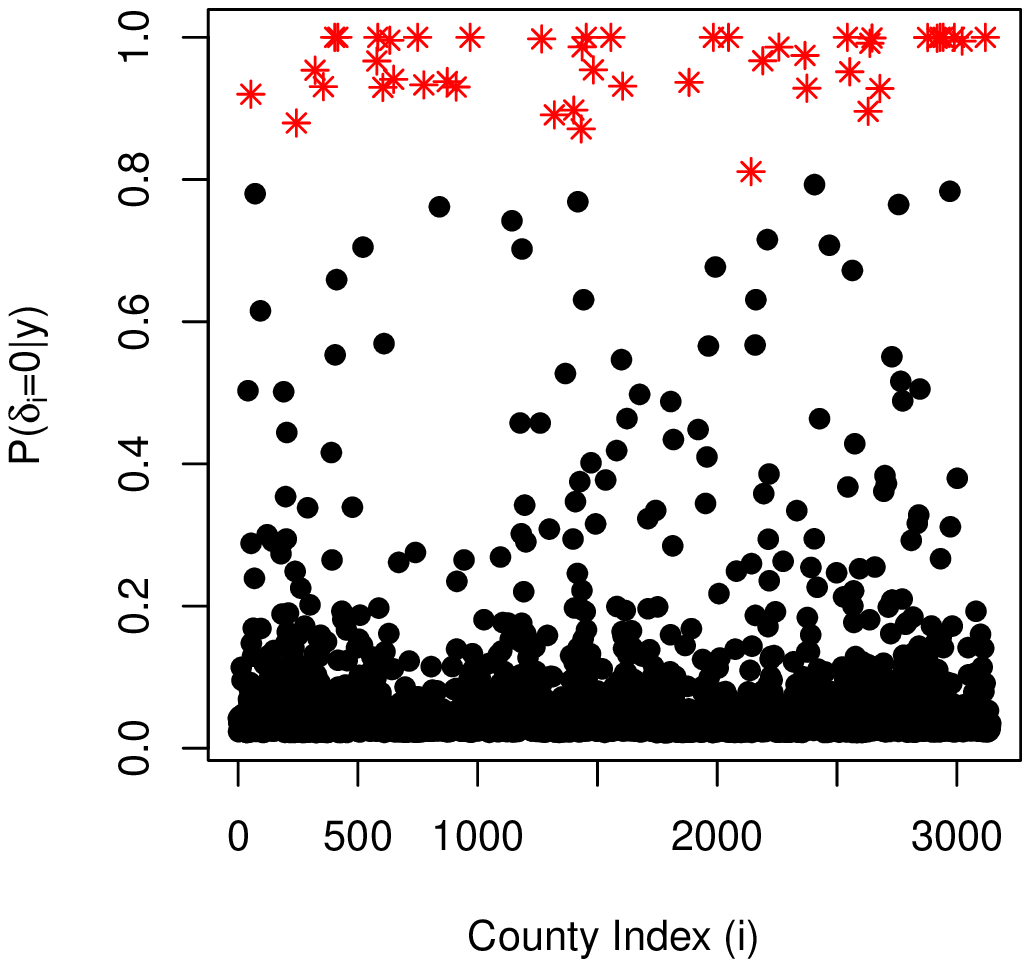} &
        \includegraphics[width=2in,height=3.0in]{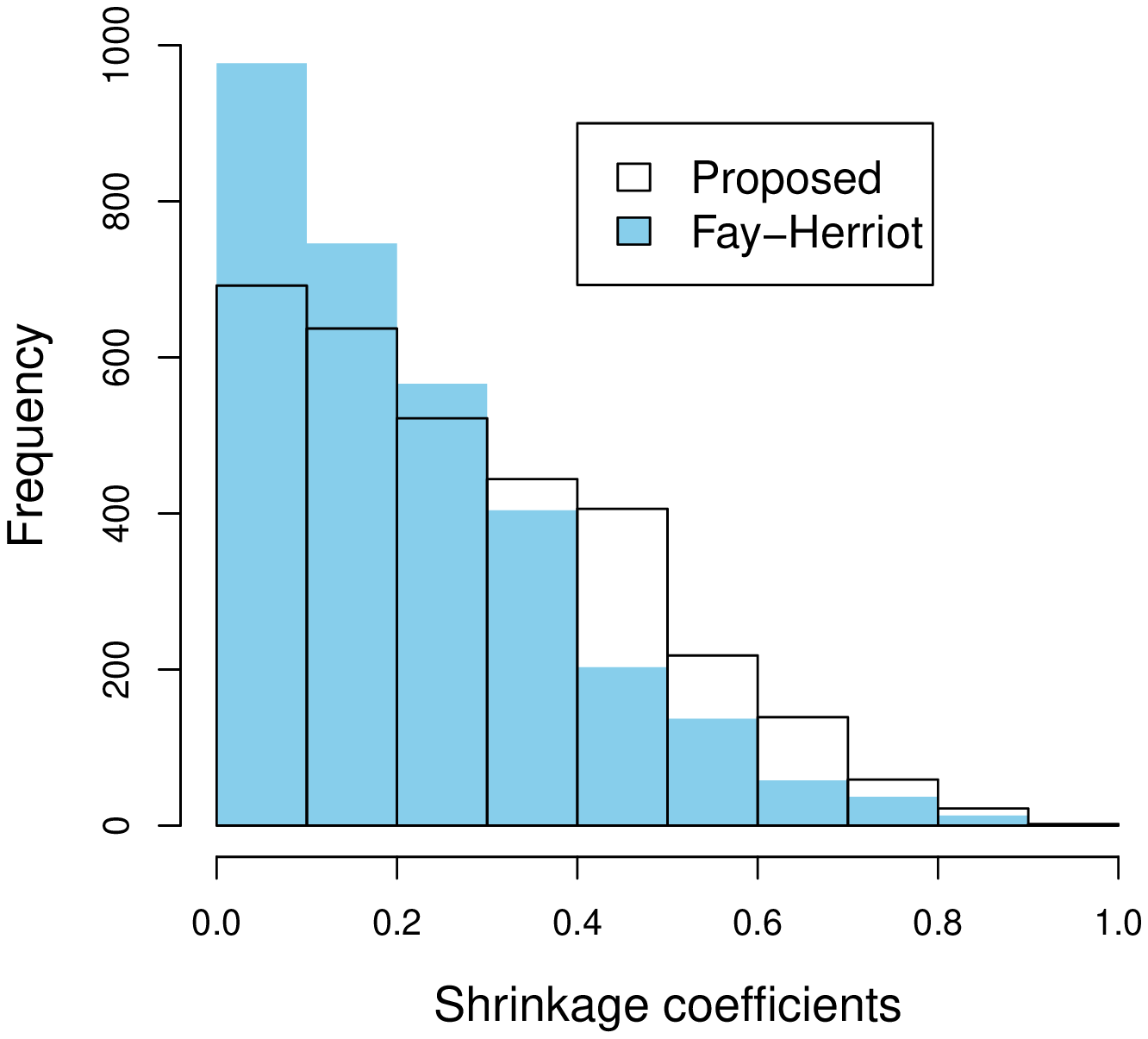} &
        \hspace{-.5in} \includegraphics[width=1.5in,height=3.0in]{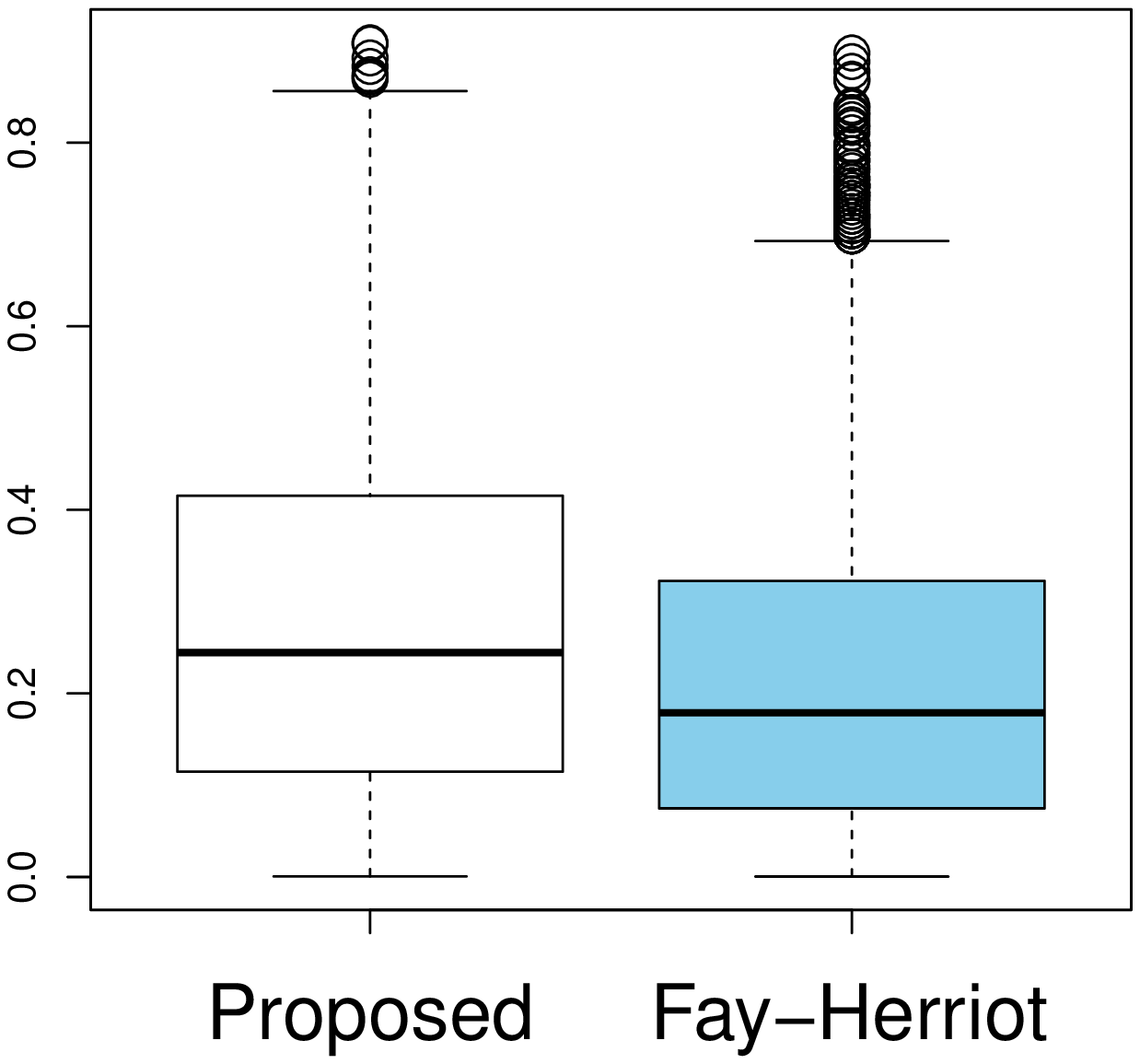}\\
        Bayes estimates of $P(\delta_{i}=1|y)$ & \multicolumn{2}{c}{Shrinkage coefficients}\\
  \end{tabular}
        \caption{Analysis of the American Community Survey data}\label{fig:box-outlier}
\end{center}
\end{figure}

\vskip -.5in

\begin{table}[h]
\begin{center}
\caption{HB estimates of model parameters (for the ACS county level poverty rates data)}\label{tab:ch21}
{\footnotesize
\begin{tabular}{crrrrr}
  \hline
                      &  Posterior               & Posterior              &\multicolumn{3}{c}{Posterior Quantiles}                         \\
\cline{4-6}
Parameter             & \multicolumn{1}{c}{Mean} & \multicolumn{1}{c}{sd} & $2.5\%$            & Median              & $97.5\%$            \\
%                      & ($\times 10^{-2}$)       & ($\times 10^{-2}$)     & ($\times 10^{-2}$) & ($\times 10^{-2}$)  & ($\times 10^{-2}$)  \\
  \hline

 $\beta_{1}$    &    0.0465        & 0.0013         &  0.0440      &  0.0465       &  0.0491    \\
 $\beta_{2}$    &    0.6605        & 0.0075         &  0.6459      &  0.6607       &  0.6748    \\
  $A_{1}$       &    0.00054       & 0.00003        &  0.00049     &  0.00054      &  0.00059   \\
  $A_{2}$       &    0.00619       & 0.00103        &  0.00454     &  0.00609      &  0.00854   \\
  $p$           &    0.0725        & 0.0237         &  0.0470      &  0.0704       &  0.1037    \\
   \hline
\end{tabular}}
\end{center}
\end{table}

\bigskip \noindent We apply our proposed method to this data set and report the results in Table~\ref{tab:ch21}. Our choices of $\alpha_{1}$ and $\alpha_{2}$ are $0.3$ and $1.3$ respectively. We have also performed further analysis with other choices of $\alpha_{1}$ and $\alpha_{2}$ within the feasible range, but results were not considerably different.  From Table~\ref{tab:ch21}, we see that the posterior mean of ${A}_{2}$(= 0.00619) is almost ten times larger than that of ${A}_{1}$(= 0.00054). In addition, the estimate $\hat{p}= 0.07$ indicates that there are about $7\%$ small areas which have much larger area specific variability compared to the most. The outlying areas can be identified by computing the Bayes estimates of posterior probabilities $P(\delta_{i}=1|y)$. We plot the estimates of these probabilities for each area in Figure~\ref{fig:box-outlier}. It shows that although most areas have low probabilities of having high random effects, some of them have higher chances of having a large variability of the model error or random small area effects. According to our analysis, approximately $7\%$ ($221$ out of $3141$) small areas have posterior probability $P(\delta_{i}=1|y)>0.15$, and approximately $1.3\%$ ($40$ out of $3141$) small areas have posterior probability $P(\delta_{i}=1|y)>0.9$.

\section{Exploration of the shrinkage coefficients}\label{sec:ch2sim}

\begin{figure}[h]
\begin{center}
\begin{tabular}{ccccc}
  \includegraphics[height=5.35cm,width=3cm]{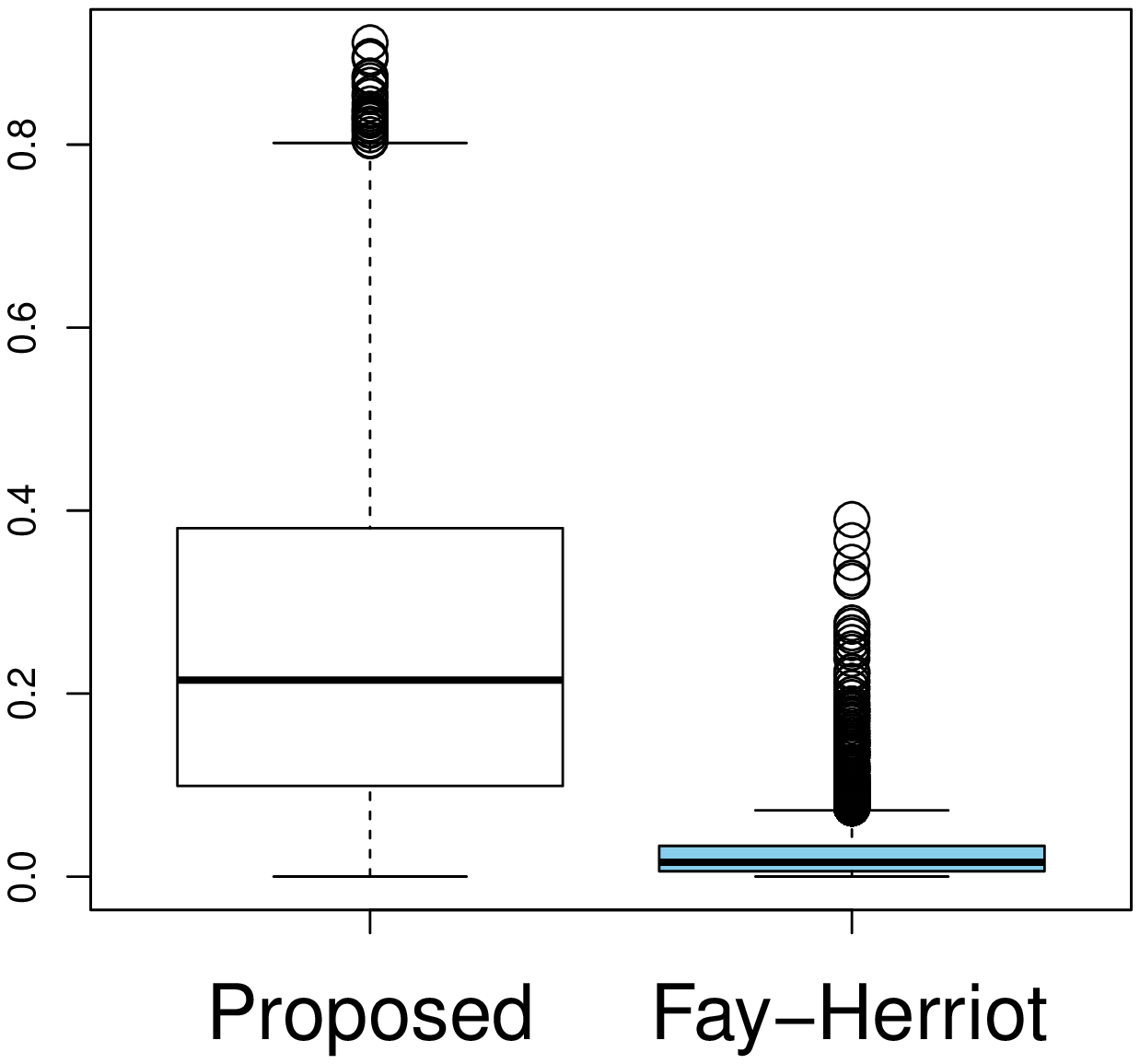} & \includegraphics[height=5.35cm,width=3cm]{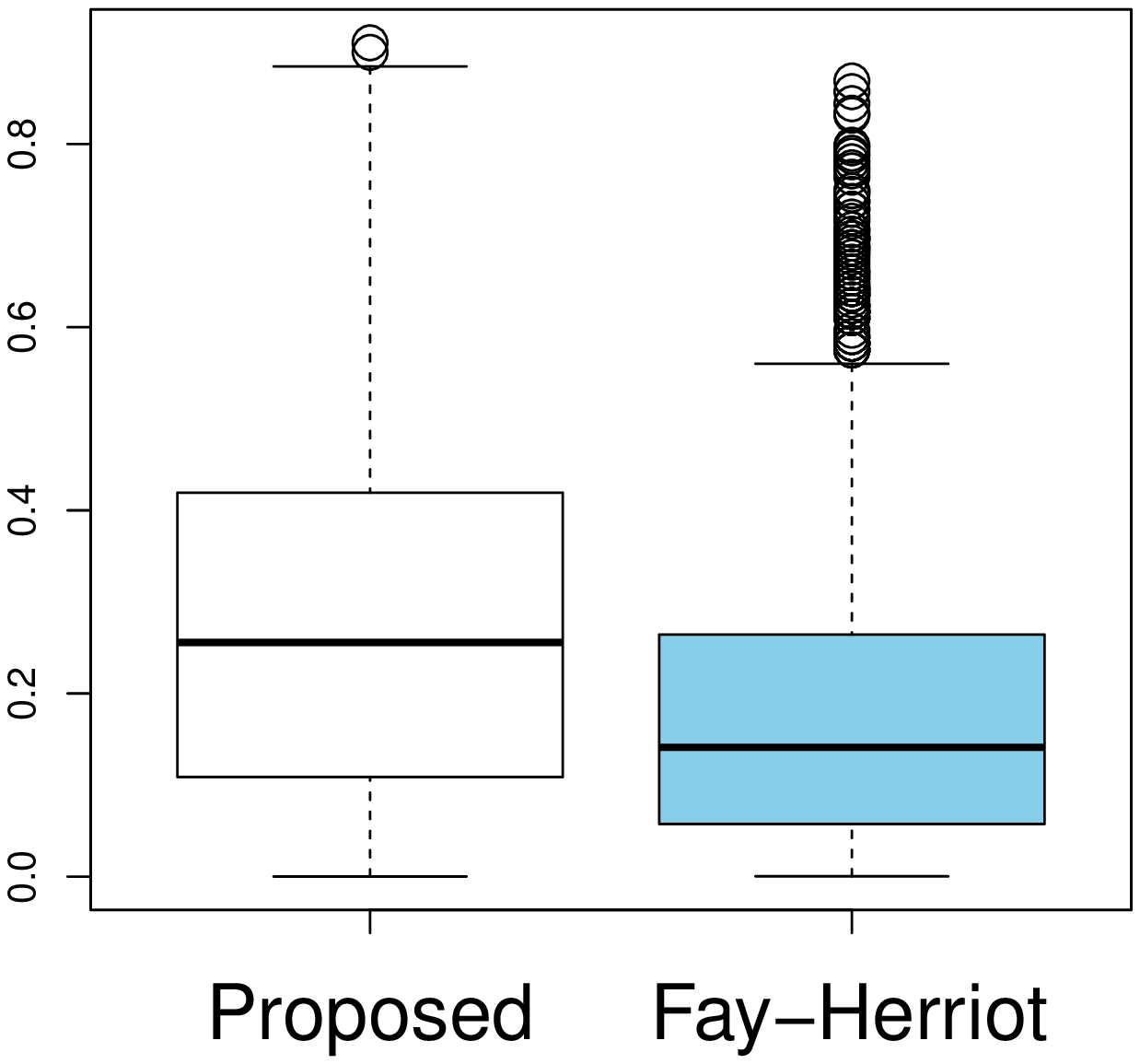} &
    \includegraphics[height=5.35cm,width=3cm]{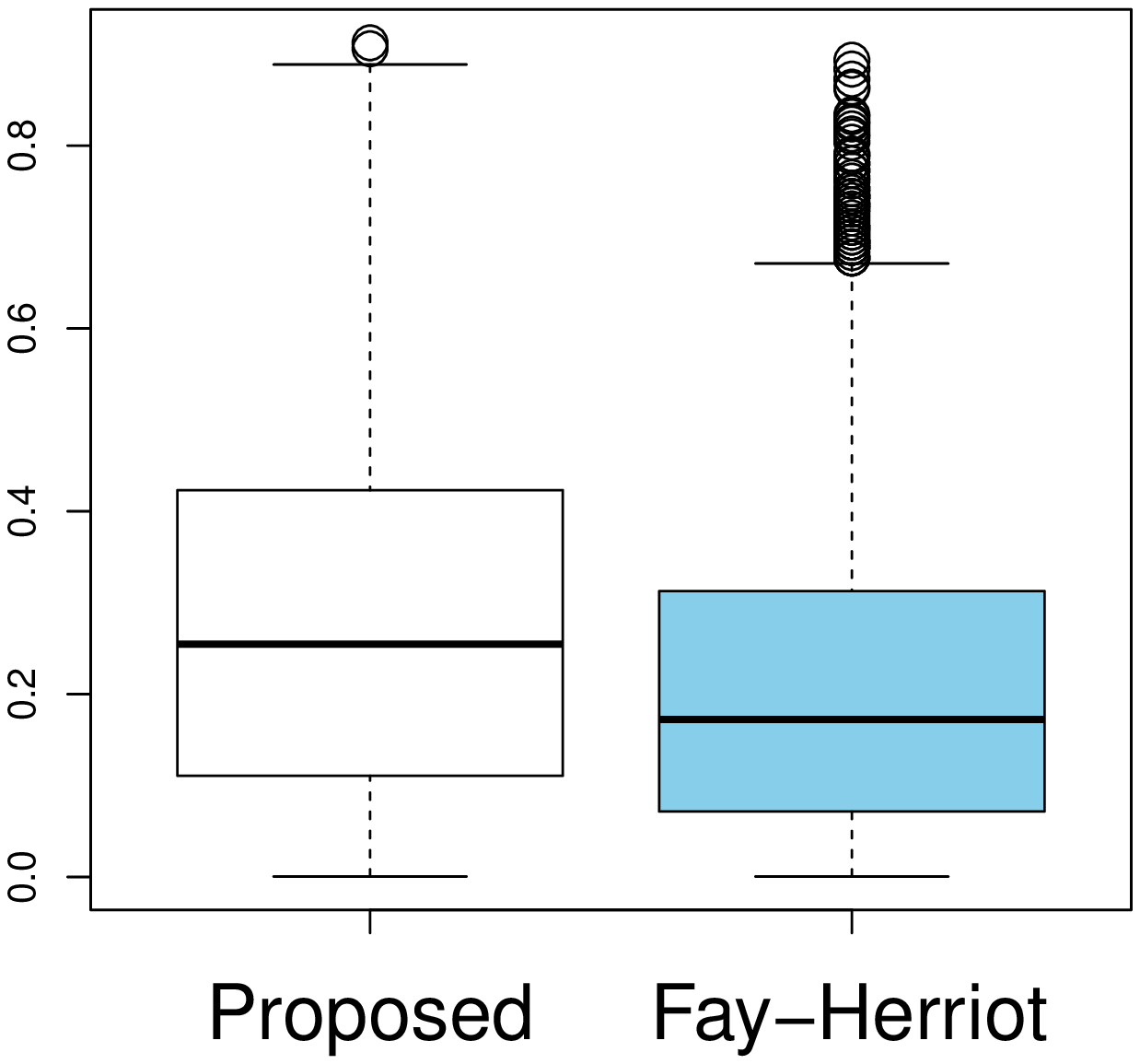} & \includegraphics[height=5.35cm,width=3cm]{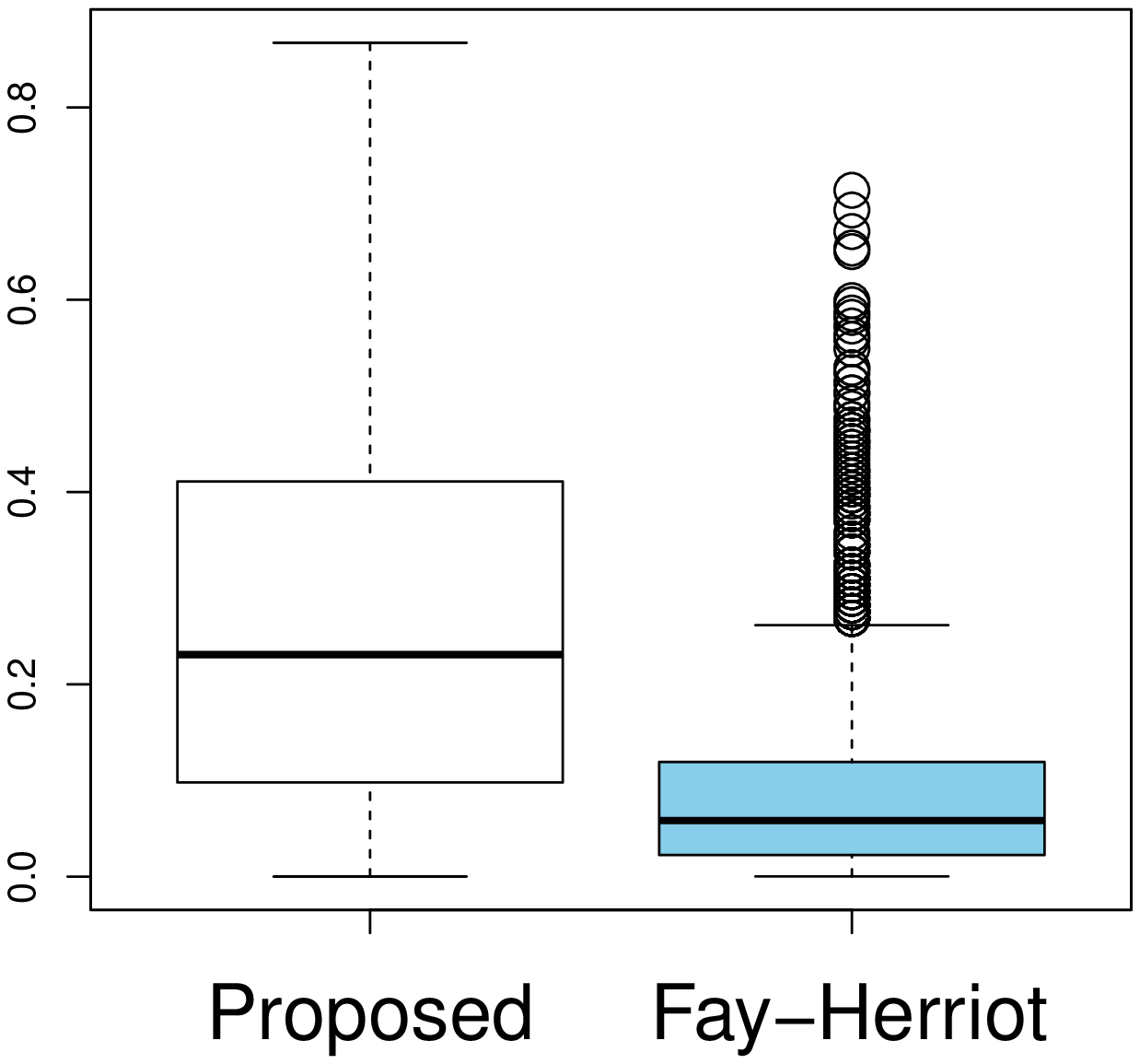} & \includegraphics[height=5.35cm,width=3cm]{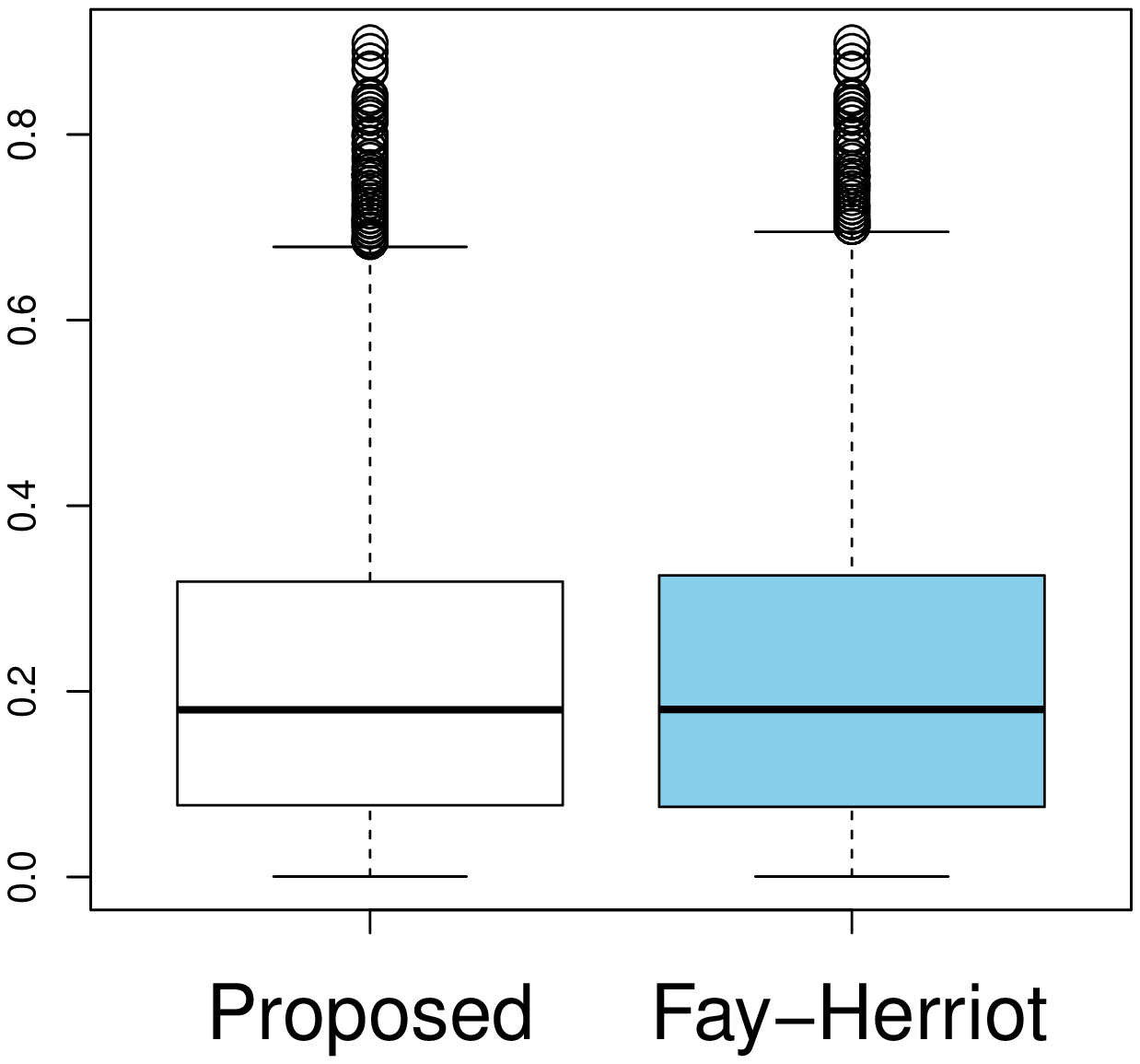} \\
    (a) & (b) &
   (c) & (d) & (e)
  \end{tabular}
\caption{ \textit{Boxplots of estimated shrinkage coefficients for two methods. In plots (a)-(d), data are partially simulated for some small areas by drawing random effects from  (a) $t_{1},$ (b) $t_{2},$ (c) $t_{3},$ (each of (a)-(c) scale adjusted) and (d) $N(0,5^2\times({0.03})^2)$.  In the plot (e), we fully simulate the data for all areas by drawing random effects from $N(0,({0.03})^2)$.}}\label{fig:shrinkage_boxplot}
\end{center}
\end{figure}

We compare the shrinkage coefficients resulting from our proposed method with those resulting  from the standard Fay Herriot model. By simulations we demonstrate that our proposed method usually provides better shrinkage than the Fay-Herriot method in presence of outliers in the data. On the other hand, simulated data from a standard Fay-Herriot model yield shrinkage coefficients based on the proposed model that are very similar to those based on the Fay-Herriot model. These two simulations, presented in Figure~\ref{fig:shrinkage_boxplot} essentially show the robustness of the proposed method to outliers.

\bigskip\noindent We mentioned in Section~\ref{sec:ch2model} that the proposed method is expected to provide better overall shrinkage than Fay-Herriot method in presence of outliers. In order to demonstrate this property of the model, we conduct the following simulations. We replace the direct estimates of first 10\% small areas of the  data by simulated values and retain the rest of the data set intact. The purpose is to artificially contaminate the data set. We generate the direct estimates  of first 10\% small areas from model (\ref{model:eq1}). We use the sampling variances of these areas to generate the corresponding sampling errors. We use the estimated regression parameters $\beta=(0.06,0.6)^{T}$ and model variance $0.0009$ obtained from Fay-Herriot analysis of the original data using the Prasad-Rao method. We use these model parameter values and the values of the auxiliary variables from these 10\% small areas to retain the mean structure and variability of the small area means which are nearly similar to the original population. We introduce outliers through use of heavy tail distribution or large model variance for random effects. Random small area effects are generated from (a) $v_{i}\sim t_{1}$, (b) $v_{i}\sim t_{2}$, (c) $v_{i}\sim t_{3}$, with proper scaling for each and (d) $v_{i}\sim N(0,5^2\times a^2)$. Note that $t_{1}$ distribution is the Cauchy distribution which does not have a variance (indeed it does not have a mean either). We rescale the draws from $t_{1}$, $t_{2}$ and $t_{3}$ by multiplying by the adjusting factor, $\dfrac{N_{0.75}}{T^{df}_{0.75}}a$, where $N_{0.75}$ and $T^{df}_{0.75}$ are the $75^{th}$ percentile of $N(0,1^2)$ and $t$ (for a specified df) respectively. By multiplying this adjusting factor, we intend to match the inter-quartile range of draws from the $t$-distribution to the inter-quartile range of a $N(0,a^2)$ distribution. Since the Prasad-Rao estimate of the random effects variance based on the original data is $0.0009$, we choose $a^2=0.0009$ in order to maintain consistency.

\begin{figure}[h]
\begin{center}
\begin{tabular}{cc}
  \includegraphics[scale=.45]{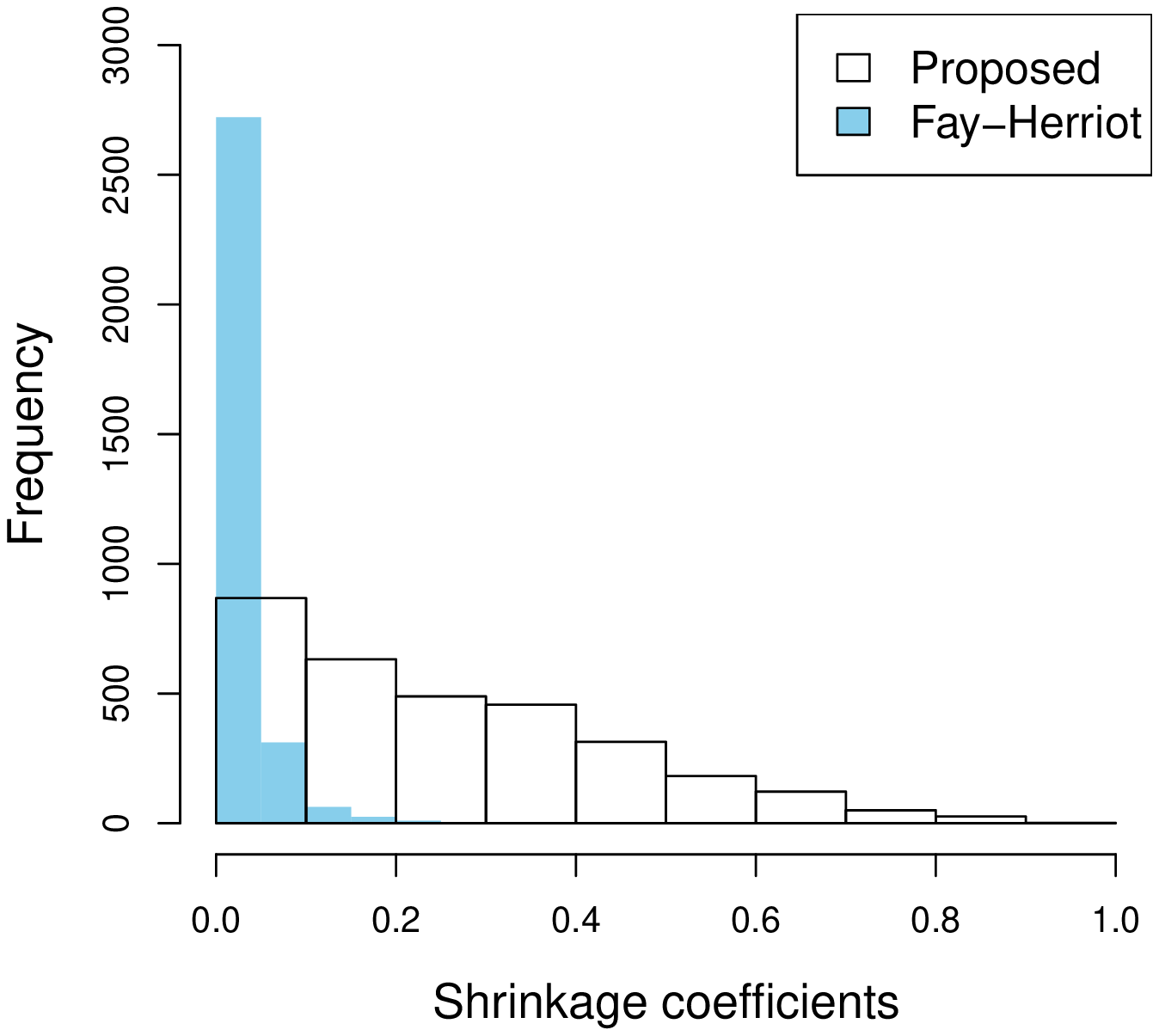} & \includegraphics[scale=.45]{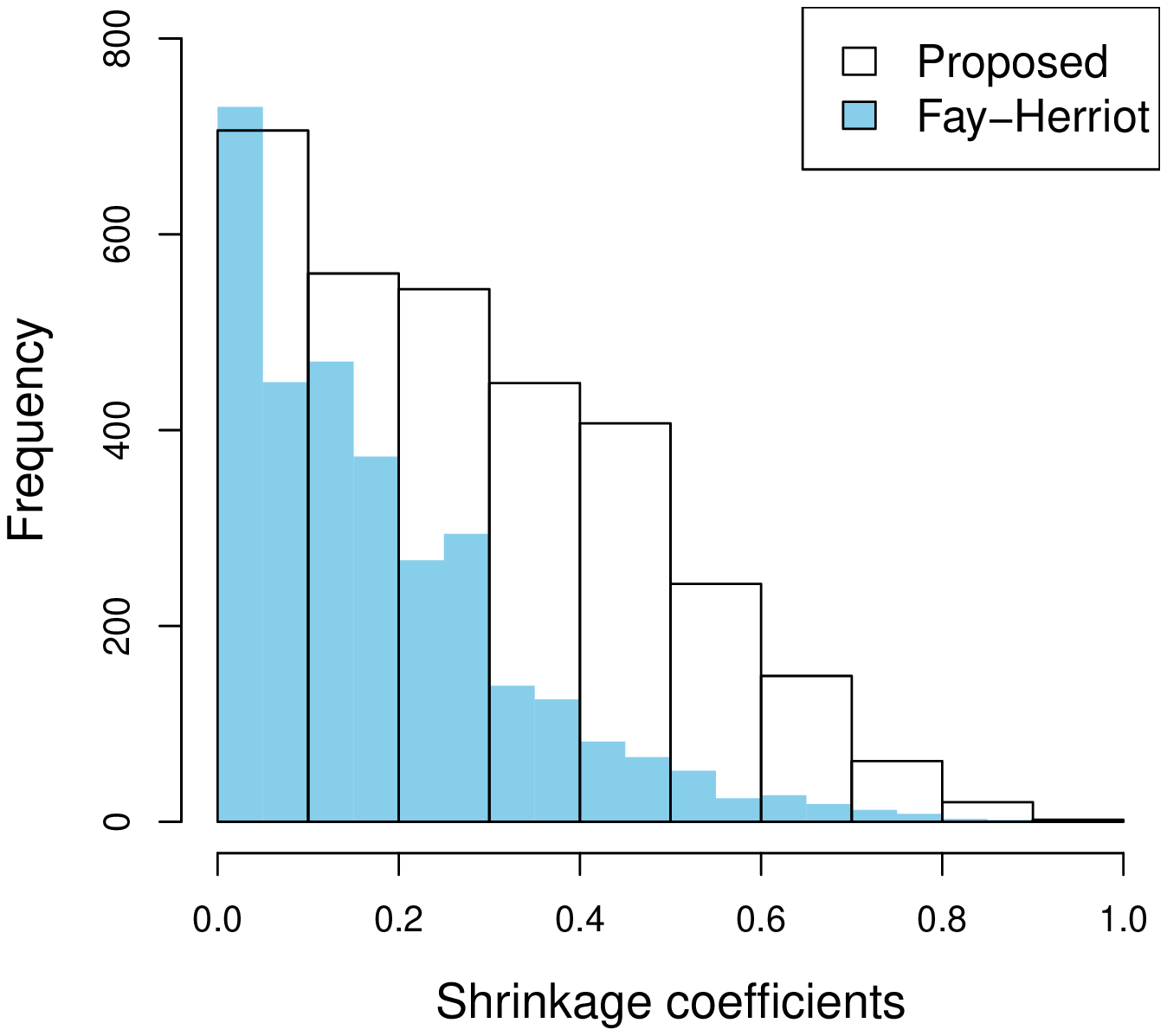} \\
    (a) & (b)\\
    \includegraphics[scale=.45]{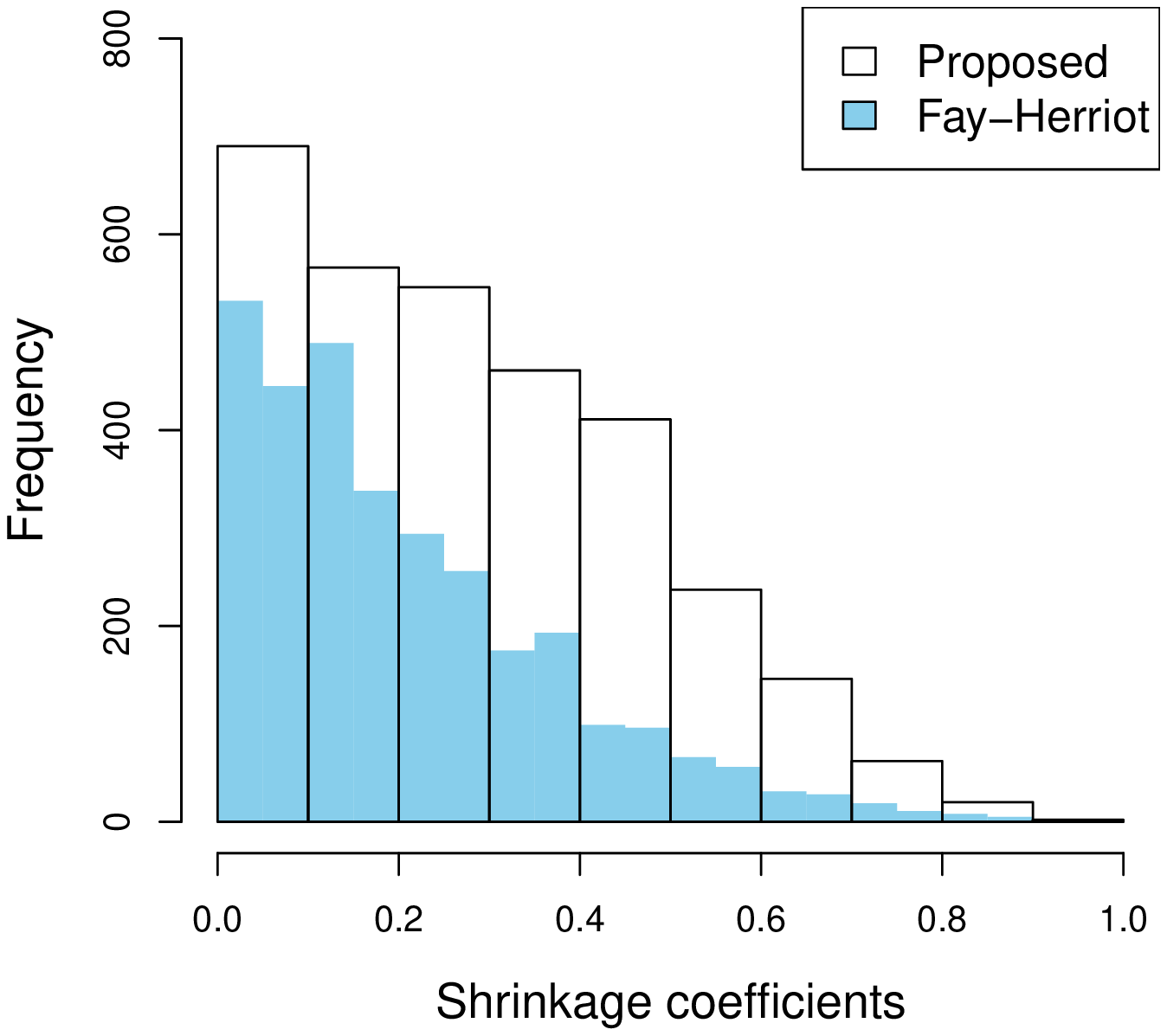} & \includegraphics[scale=.45]{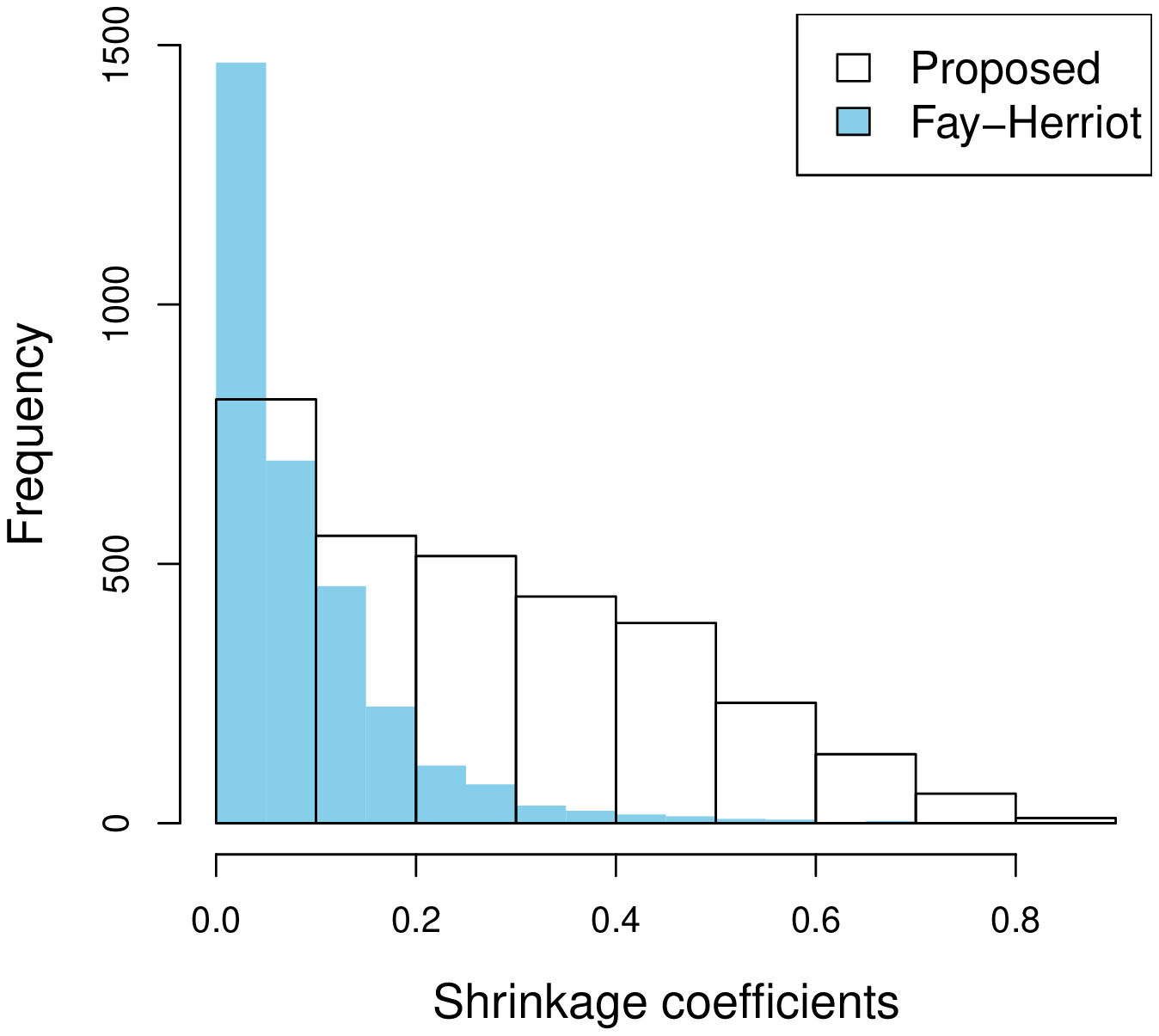}\\
    (c) & (d)
  \end{tabular}
\caption{ \textit{Histograms of estimated shrinkage coefficients of two methods when the data are partially simulated by drawing random effects from (a) $t_{1},$ (b) $t_{2},$ (c) $t_{3}$ (each of (a)-(c) scale adjusted), and (d) $N(0,5^2\times({0.03})^2)$}}\label{fig:shrinkage_hist}
\end{center}
\end{figure}

\vskip -.3in
\noindent We apply the proposed method as well as the Fay-Herriot method and compare the estimates of shrinkage coefficients in Figures~\ref{fig:shrinkage_boxplot} and~\ref{fig:shrinkage_hist}. We see from Figure~\ref{fig:shrinkage_hist} that when we partially contaminate the data set using (a) re-scaled $t_{1}$ (Cauchy) and (d) N$(0,5^2\times({0.03})^2)$, the overall shrinkage obtained from the proposed model is  considerably higher than the overall shrinkage obtained from the regular Fay-Herriot method. This result shows the flexibility of the proposed model in borrowing information from other areas when outliers in the random effects are present. Panels (b), (c) and (e) of Figure~\ref{fig:shrinkage_boxplot} show that the proposed method performs similarly as the Fay-Herriot method when the departure of the random effects distribution from the normal is moderate or none.

\clearpage

\section{Performance of the proposed method}\label{sec:performance}
\noindent In order to evaluate the performance of the proposed model described in Section~\ref{sec:ch2model}, we conduct a simulation study. This analysis is based on simulated data sets generated under different settings. For each $m=100$, $500$ and $1000$, we generated $100$ data sets. Here we set $r=2$,  $x=(1,x_{1})^{T}$ and generate $m$ copies of $x_{1}$ from N$(10,(\sqrt{2})^{2})$. For each choice of $m$, the set of covariates is  generated exactly once and used the set for all $100$ data sets.  Our choice of $\beta$ is $\beta=(20,1)^{T}$. The sampling error $e_{i}$'s are generated from $N(0,D_{i})$, $i=1,\dots,m$, where $D_{i}$'s are from the set $\{0.5$, $1$, $1.5$, $2$, $2.5$, $3$, $3.5$, $4$, $4.5$, $5\}$, each value in the set is allocated to the same number of small areas. Random effects in model (\ref{model:eq1}) are generated under three different settings,
\begin{align}
v_{i} &\sim N(0,1^2), \label{eqn:normal}\\
v_{i} &\sim (1-\delta_{i}) N(0,1^2)+ \delta_{i}N(0,5^2),~~\text{and} \label{eqn:mix}\\
v_{i} &\sim t_{3}, \label{eqn:t}
\end{align}
where $i=1,\dots,m$. For the normal-mixture setup (\ref{eqn:mix}), we set $\delta_{i}=1$ for each $i$ multiple of $5$ and keep rest of the $\delta_{i}=0$, the simulated data sets contain $20$\% observations from the normal distribution with variance $25$.  Based on a generated set of of $v_i$'s , we compute both the $\theta_i$'s and $y_i$'s by (\ref{model:eq1}).  For each of $100$ simulated data sets for each setting, we predict  $\theta_{i}$'s based on Fay-Herriot model and the proposed area-level normal mixture model. We measure the performance of each prediction method by computing (empirical) mean squared error (MSE)= $\frac{1}{m}\sum\limits_{i=1}^{m} (\theta_{i}-\hat{\theta}_{i})^2$, mean absolute error (MAE)=$\frac{1}{m}\sum\limits_{i=1}^{m} |\theta_{i}-\hat{\theta}_{i}|$, mean relative squared error (MRSE)= $\frac{1}{m} \sum\limits_{i=1}^{m} \frac{(\theta_{i}-\hat{\theta}_{i})^2}{\theta^{2}_{i}}$ and  mean relative absolute error (MRAE)= $\frac{1}{m}\sum\limits_{i=1}^{m}  \frac{|\theta_{i}-\hat{\theta}_{i}|}{\theta_{i}}$, where $\theta_{i}$'s are true and $\hat{\theta}_{i}$'s are estimated small area means (for our simulation setup, all the $\theta_i$'s are positive). These empirical deviation measures are typically used in small area estimation literature to compare the accuracy of various estimation methods (Rao, 2003). For each simulated dataset, we compute MSE, MAE, MRAE and MRSE for two different methods and report the average values based on all simulated data sets. Results of the simulation study are presented in Tables~\ref{tab:allsim} and~\ref{tab:mixsim}. {In Table~\ref{tab:allsim} we report the MSE and MAE and in Figure~4 we plot the MRAE and MRSE based on the overall simulation study}. Table~\ref{tab:mixsim} shows a more detailed result when the $v_i$'s are drawn according to equation~(\ref{eqn:mix}). From Table~\ref{tab:mixsim} we can compare performance of two prediction methods for outlying areas (random effects drawn from $N(0,5^2)$) and non-outlying areas (random effects drawn from $N(0,1^2)$), separately. Simulation results indicate that the proposed method tends to perform better than the Fay-Herriot method when the possibility of presence of outliers is high, and performs similarly otherwise.

\begin{table}[h]
\begin{center}
\caption{Comparison of the methods based on simulated MSE and MAE of prediction. Results are based on $100$ simulated data sets}\label{tab:allsim}
\begin{tabular}{lllllllllll}
\hline
                                           & &  &\multicolumn{2}{c}{m=100} && \multicolumn{2}{c}{m=500} &&\multicolumn{2}{c}{m=1000}\\
\cline{4-5} \cline{7-8} \cline{10-11}
Scenario                                   &           &  & Proposed  &  FH     &&  Proposed & FH      && Proposed& FH      \\
\hline
\multirow{2}{*}{(\ref{eqn:normal})~Normal}  & MSE      &  & 0.72      & 0.71    &&   0.69    & 0.69    &&  0.68   & 0.68  \\
                                            & MAE      &  & 0.67      & 0.67    &&   0.66    & 0.66    &&  0.66   & 0.65  \\
                                           %& MRSD     &  & 0.0008    & 0.0008  &&   0.0008  & 0.0008  &&  0.0008 & 0.0008  \\
                                           %& MRAD     &  & 0.0227    & 0.0226  &&   0.0220  & 0.0220  &&  0.0219 & 0.0219  \\
\hline
\multirow{2}{*}{(\ref{eqn:mix})~Mixture}    & MSE      &  & 1.48      & 1.75    &&   1.49    & 1.81    &&  1.30   & 1.87  \\
                                            & MAE      &  & 0.86      & 1.01    &&   0.85    & 0.98    &&  0.84   & 1.04  \\
                                           %& MRSD     &  & 0.0018    & 0.0021  &&   0.0018  & 0.0021  &&  0.0016 & 0.0024  \\
                                           %& MRAD     &  & 0.0296    & 0.0346  &&   0.0287  & 0.0327  &&  0.0287 & 0.0357  \\
\hline
\multirow{2}{*}{(\ref{eqn:t})~$t_{3}$}      & MSE      &  & 1.14      & 1.27    &&   1.01    & 1.20    &&  1.14   & 1.30  \\
                                            & MAE      &  & 0.83      & 0.84    &&   0.79    & 0.81    &&  0.80   & 0.84  \\
                                           %& MRSD     &  & 0.0014    & 0.0015  &&   0.0013  & 0.0014  &&  0.0013 & 0.0022  \\
                                           %& MRAD     &  & 0.0280    & 0.0293  &&   0.0263  & 0.0276  &&  0.0272 & 0.0290  \\
                            \hline
\end{tabular}
\end{center}
\end{table}

\vskip -.3in

\begin{figure}[h]
\begin{center}
\begin{tabular}{cc}
  \includegraphics[scale=.5]{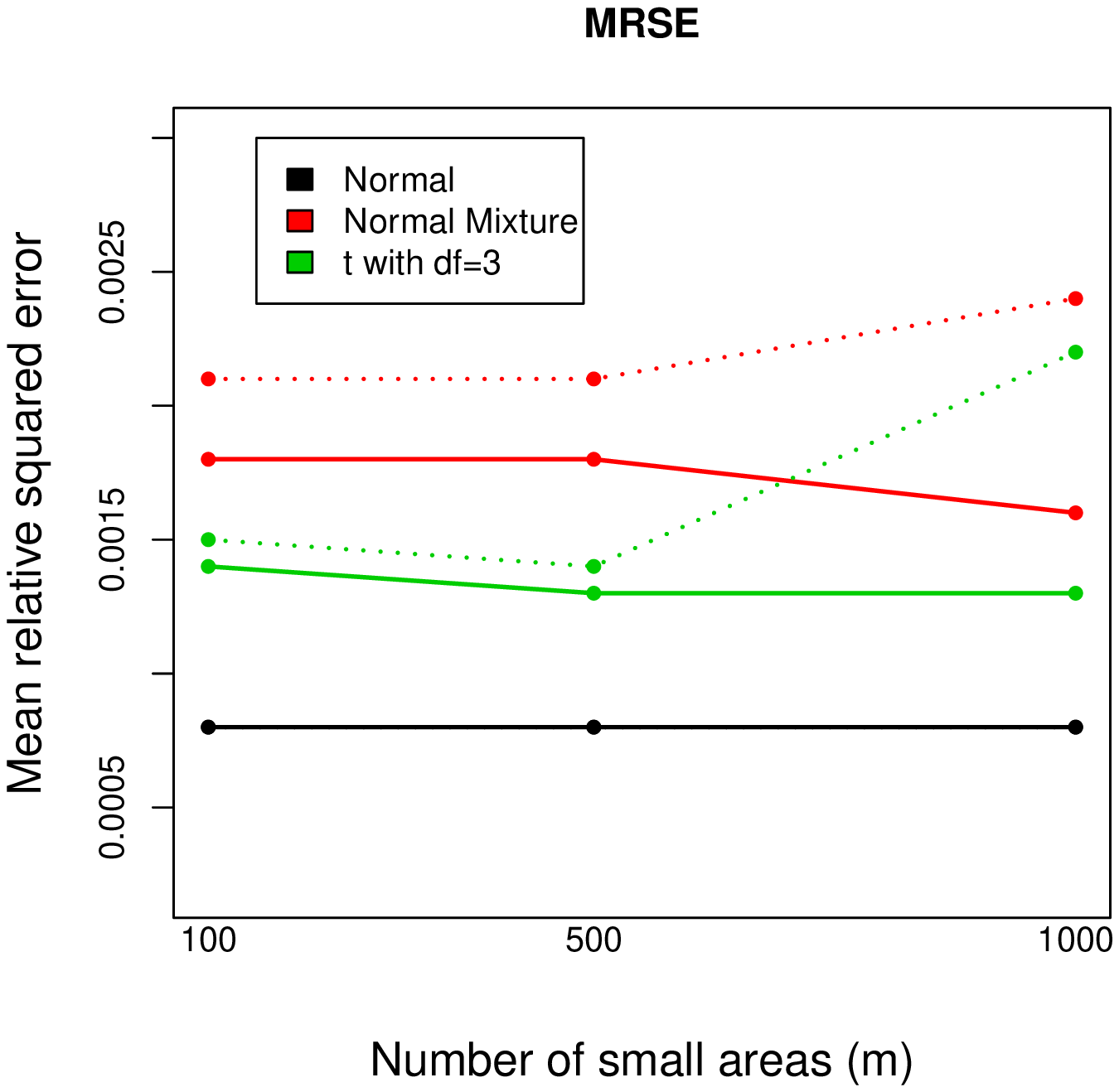} & \includegraphics[scale=.5]{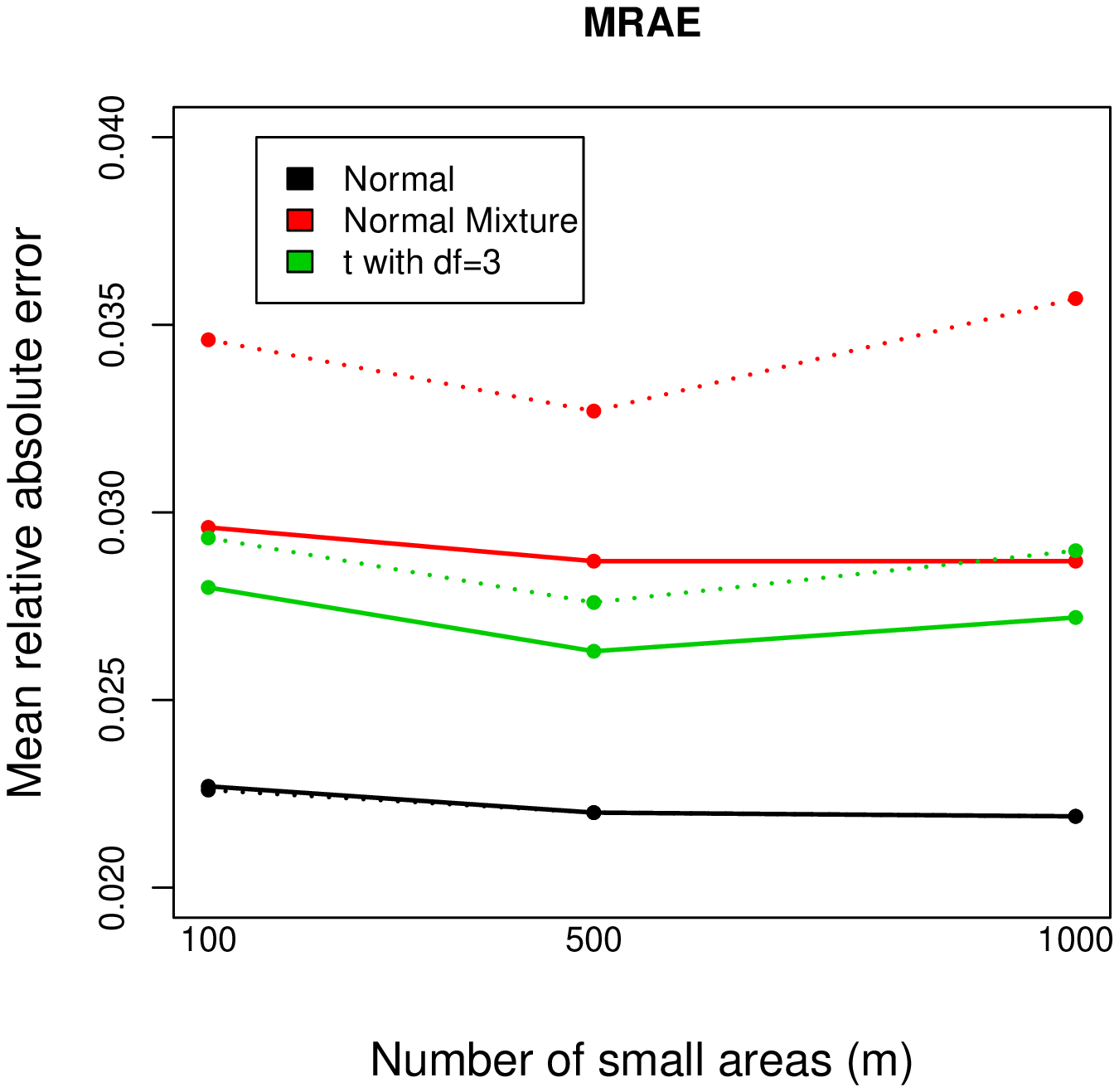} \\
    (a) & (b)\\
  \end{tabular}
\caption{ \textit{(a) Mean relative squared error (MRSE) and (b) mean relative absolute error (MRAE) based on $100$ simulated data sets; Dotted line for Fay-Herriot method and  solid line for the proposed method}}\label{fig:simul}
\end{center}
\end{figure}

\clearpage

\begin{table}[h]
\begin{center}
\caption{Comparison of the methods based on simulated MSE, MAE, MRSE and MRAE of prediction. Results are based on $100$ simulated data sets. Performance of the methods are compared separately for outlying and non-outlying areas based on the simulation design.}\label{tab:mixsim}
\vspace{.1in}
\begin{tabular}{lllllllllll}
\hline
&&\multicolumn{9}{c}{Scenario (\ref{eqn:mix})~Mixture}\\
                                           & &  &\multicolumn{2}{c}{m=100} && \multicolumn{2}{c}{m=500} &&\multicolumn{2}{c}{m=1000}\\
\cline{4-5} \cline{7-8} \cline{10-11}
                                       &&                & Proposed  &  FH     &&Proposed& FH      && Proposed& FH     \\
\hline

\multirow{8}{*}{}   & \multirow{2}{*}{MSE}      & $A_1 = 1^2$  & 0.90  & 1.26  && 0.80 & 1.06  &&  0.80 & 1.32 \\
                    &                           & $A_2 = 5^2$  & 3.39  & 3.69  && 4.25 & 4.80  &&  3.28 & 4.03 \\
                    \cline{2-11}
                    & \multirow{2}{*}{MAE}      & $A_1 = 1^2$  & 0.73  & 0.88  && 0.69 & 0.82  &&  0.70 & 0.91 \\
                    &                           & $A_2 = 5^2$  & 1.43  & 1.47  && 1.49 & 1.61  &&  1.39 & 1.59 \\
                    \cline{2-11}
                    & \multirow{2}{*}{100$\times$MRSE} & $A_1 = 1^2$  & 0.10  & 0.14  && 0.09 & 0.12  &&  0.09 & 0.15 \\
                    &                           & $A_2 = 5^2$  & 0.43  & 0.50  && 0.53 & 0.56  &&  0.44 & 0.61 \\
                    \cline{2-11}
                    & \multirow{2}{*}{10$\times$MRAE}  & $A_1 = 1^2$  & 0.25  & 0.30  && 0.23 & 0.27  &&  0.24 & 0.30 \\
                    &                           & $A_2 = 5^2$  & 0.50  & 0.52  && 0.51 & 0.54  &&  0.49 & 0.57 \\
                            \hline

                            \hline
\end{tabular}
\end{center}
\end{table}

%\clearpage

\section{Discussion}\label{Discussion}

\bigskip\noindent In this paper, we propose a robust alternative to Fay-Herriot model. The proposed hierarchical Bayesian estimation procedure is straightforward. Other robust alternative is a $t$-distribution for the random effects, which requires information regarding the degrees of freedom. Xie et al. (2007) proposed a method to estimate degrees of freedom, {however, Bell and Huang pointed out that only a very limited information can be extracted from the data regarding degrees of freedom parameter}. We propose a method based on noninformative priors for the {parameters}. We provide sufficient conditions for the propriety of the resulting posterior distributions.

\bigskip\noindent Model-based small area estimates depend on the accuracy of the underlying model assumptions. Larger values of area specific random effects may be caused by poor choice of the linking model or lack of predictive quality of the auxiliary variables. If the model-based estimates of area specific random effects are significantly larger for some areas compared to the other areas, it is probably meaningful to retain the direct estimates instead of model-based estimates for those areas to avoid possible inaccuracy. Although we should be cautious in this recommendation if there is any indication that the sampling variance is underestimated.

\bigskip\noindent Datta and Lahiri (1995) recommended heavy-tailed priors for random effects by emphasizing the fact that estimators obtained by using these priors are similar to direct estimators for the areas with extreme observations. However, the estimators for non-outlying areas should shrink direct estimators more toward synthetic estimators and the magnitude of this shrinkage may depend on the quality of the auxiliary information. While for an outlying observation our model limits shrinkage of  Bayes predictor to the synthetic estimator, for non-outlying observations it enables the Bayes predictors to retain shrinkage to the synthetic estimator when the regression model provides a good fit.

\subsection*{Acknowledgment}

\noindent Authors are thankful to Dr. Jerry Maples of Census Bureau for providing and explaining the poverty data in our application. They are also grateful to him and to Dr. William R. Bell for an internal review of an earlier version of the manuscript. Their many valuable comments, particularly some substantive comments of Dr. Bell, led to a significantly improved manuscript. Research of Mandal is partially supported by the NSA Grant H98230-13-1-0251.

\section*{References}

\begin{enumerate}

\item Angers, J. F., and Berger, J. O. (1991). Robust hierarchical Bayes estimation of exchangeable means. {\it Canadian Journal of Statistics}, {\bf 19}, 39-56.

\item Bell, W. R., and Huang, E. T.  (2006). Using $t$-distribution to deal with outliers in small area estimation. {\it Proceedings of Statistics Canada Synposium 2006 Methodological issues in measuring population health}.

\item Berger, J. O. (1984). The robust Bayesian viewpoint (with discussion). In {\it Robustness of Bayesian Analysis}, Ed.: J. Kadane, North-Holland, Amsterdam.

\item Chambers, R., Chandra, H., Salvati, N., and Tzavidis, N. (2014). Outlier robust small area estimation. {\it Journal of the Royal Statistical Society, Series B:}, {\bf 76}, 47$-$69.

\item Datta, G. S. (2009). Model-based approach to small area estimation. {\it Handbook of Statistics: Sample Surveys: Inference and Analysis}, {\it 29B}, Eds.: D. Pfeffermann and C.R.
Rao, The Netherlands: North-Holland. 251-288.

\item  Datta, G. S., and Ghosh, M. (2012). Small area shrinkage estimation. {\it Statistical Science}, {\bf 27}, 95-114.

\item Datta, G. S., Hall, P. G., and Mandal, A. (2011). Model selection by testing for the presence
of small-area effects in area-level data. {\it Journal of the American Statistical Association}, {\bf 106}, 362-374.

\item  Datta, G. S., and Lahiri, P. (1995). Robust hierarchical Bayesian estimation of small area characteristics in presence of covariates and outliers. {\it Journal of Multivariate Analysis}, {\bf 54}, 310-328.

\item Datta, G. S., and Mandal, A. (2015). Small area estimation with uncertain random effects. To appear in {\it Journal of the American Statistical Association}, {\bf 110}, DOI: 10.1080/01621459.2015.1016526.

\item Datta, G. S., Rao, J. N. K., and Smith, D. (2005). On measuring the variability of small area estimators under a basic area level model. {\it Biometrika}, {\bf 92,} 183-196.

\item Dey, D. K., and Berger, J. O. (1983). On truncation of shrinkage estimators in simultaneous estimation of normal means.
{\it Journal of the American Statistical Association}, {\bf 78}, 865-869.

\item Efron, B., and Morris, C. (1971). Limiting the risk of Bayes and empirical Bayes estimators, Part I: The Bayes Case. {\it Journal of the American Statistical Association}, {\bf 67}, 130-139.

\item  Fay, R. E., and Herriot, R. A. (1979). Estimates of income for small places: an application of James-Stein procedures to census data. {\it Journal of the American Statistical Association}, {\bf 74}, 269-277.

\item  Ghosh, M., and Rao, J. N. K. (1994). Small are estimation: an appraisal. {\it Statistical Science}, {\bf 9}, 55-93.

\item Ghosh, M., Maiti, T., and Roy, A. (2008). Influence functions and robust Bayes and empirical Bayes small area estimation. {\it Biometrika}, {\bf 95}, 573$-$585.

\item Jiang, J., and Lahiri, P. (2006). Mixed model prediction and small area estimation. {\it Test},  {\bf 15}, 1-96.

\item Pfeffermann, D. (2013). New important developments in small area estimation. {\it Statistical Science},  {\bf 28}, 40-68.

\item Prasad, N. G. N., and Rao, J. N. K. (1990). The estimation of the mean squared error of small-area estimators.  {\it Journal of the American Statistical Association}, {\bf 85}, 163-171.

\item Rao, J. N. K. (2003). {\it Small Area Estimation}. Wiley-Interscience, Hoboken, NJ.

\item Rao, J. N. K. (2011). Impact of frequentist and Bayesian methods on survey sampling practice: a selective appraisal.  {\it Statistical Science}, {\bf 26}, 240-256.

\item Scott, J. G., and Berger, J. O. (2006). An exploration of aspects of Bayesian multiple testing. {\it Journal of Statistical Planning and Inference}, {\bf 136}, 2144-2162.

\item Sinha, S. K., and  Rao, J. N. K. (2009). Robust small area estimation. {\it The Canadian Journal of Statistics}, {\bf 37}, 381-399.

\item Stein, C. M. (1956). Inadmissibility of the usual estimator for the mean of a multivariate normal distribution. { \it Proceedings of the Third Berkeley Symposium on Mathematical Statistics and Probability} 1954$-$1955, {\bf I}, 197-206. Univ. California Press, Berkeley.

\item Stein, C. M. (1981). Estimation of the mean of a multivariate normal distribution. {\it Annals of Statistics}, {\bf 9}, 1135-1151.

\item  Xie, D., Raghunathan, T. E., and Lepkowski, J. M. (2007). Estimation of the proportion of overweight individuals in small areas--a robust extension of the Fay-Herriot model. {\it Statistics in Medicine}, {\bf 26}, 2699-2715.

\end{enumerate}

%\newpage
\section*{Appendix}

\subsection*{Gibbs sampling for the proposed model}\label{sec:computation}
In order to apply our model, we use Gibbs sampling. We derive the set of full conditional distributions from the posterior joint density of $\theta=(\theta_{1},\dots,\theta_{m})^{T}$, $\beta=(\beta_{1},\dots,\beta_{r})^{T}$, $\delta=(\delta_{1},\dots,\delta_{m})^{T}$, $A_{1}$, $A_{2}$ and $p$, which is given by
\begin{align}
\label{eqn:ch2joint}
\pi(\theta,\beta,A_{1},A_{2},\delta,p|y) & \propto \left\{\prod_{i=1}^{m}\exp\left\{ -\frac{(y_{i}-\theta_{i})^2}{2D_{i}}\right\} \right\} \prod_{i=1}^{m}\bigg[ \left\{ \dfrac{1}{\sqrt{A_{1}}}\times \exp\left\{-\frac{(\theta_{i}-x^{T}_{i}\beta)^2}{2A_{1}} \right\}\right\}^{\delta_{i}} \nonumber\\
&~~~~~\times \left\{ \dfrac{1}{\sqrt{A_{2}}}\times \exp\left\{-\frac{(\theta_{i}-x^{T}_{i}\beta)^2}{2A_{2}} \right\}\right\}^{1-\delta_{i}} p^{\delta_{i}}(1-p)^{1-\delta_{i}} \bigg] \nonumber \\
&~~~~~~~~~\times A^{-\alpha_{1}}_{1}A^{-\alpha_{2}}_{2}\times I(0<A_{1}<A_{2}).
\end{align}
From (\ref{eqn:ch2joint}), we get the following full conditional distributions:
\begin{enumerate}[(I)]
\item $ \theta_{i}|\beta,A_{1},A_{2},\delta,p,y \stackrel{\text{ind}}{\sim}
 N\Big(\dfrac{D_{i}x^{T}_{i}\beta +A_{2-\delta_{i}}y_i }{D_{i}+A_{2-\delta_{i}}}$ , $\dfrac{D_{i}A_{2-\delta_{i}}}{D_{i}+A_{2-\delta_{i}}}\Big)$,
    $i=1,\dots,m$;
\item $\beta|\theta,A_{1},A_{2},\delta,p,y \sim N\bigg( \left[ \sum\limits_{i=1}^{m} A_{2-\delta_{i}}^{-1} x_{i}x^{T}_{i} \right]^{-1}\left[ \sum\limits_{i=1}^{m} A_{2-\delta_i}^{-1}x_{i}\theta_{i} \right], \left[ \sum\limits_{i=1}^{m} A_{2-\delta_{i}}^{-1} x_{i}x^{T}_{i} \right]^{-1}\bigg)$;
\item $p|\theta,\beta,A_{1},A_{2},\delta,y \sim Beta\left( \sum\limits_{i=1}^{m}\delta_{i}+1,m- \sum\limits_{i=1}^{m}\delta_{i}+1\right)$;
\item $A_{1}|A_{2},\theta,\beta,\delta,p,y$ has the pdf $f_{1}(A_{1})$, where,
\begin{center}
      $f_{1}(A_{1}) \propto A_{1}^{-(\alpha_{1}+\sum_{i=1}^{m}\frac{\delta_{i}}{2})}\exp\left\{-\sum\limits_{i=1}^m\dfrac{\delta_{i}(\theta_{i}-x^{T}_{i}\beta)^2}{2A_{1}} \right\}I(A_{1}<A_{2})$,
\end{center}
\item $A_{2}|A_{1},\theta,\beta,\delta,p,y$ has the pdf $f_{2}(A_{2})$, where,
\begin{center}
      $f_{2}(A_{2}) \propto A_{2}^{-(\alpha_{2}+\sum_{i=1}^{m}\frac{(1-\delta_{i})}{2})}\exp\left\{-\sum\limits_{i=1}^m\dfrac{(1-\delta_{i})(\theta_{i}-x^{T}_{i}\beta)^2}{2A_{2}} \right\}I(A_{1}<A_{2})$,
\end{center}
\item For $i=1,\dots,m,~$$\delta_i|\theta,\beta,A_{1},A_{2},p,y$ are independent with \\
$P(\delta_{i}=1|\theta,\beta,p,y)=\dfrac{p \times \exp\left\{-\frac{(\theta_{i}-x^{T}_{i}\beta)^2}{2A_{1}}\right\}A_{1}^{-\frac{1}{2}}}{p \times \exp\left\{-\frac{(\theta_{i}-x^{T}_{i}\beta)^2}{2A_{1}}\right\}A_{1}^{-\frac{1}{2}}+(1-p) \times \exp\left\{-\frac{(\theta_{i}-x^{T}_{i}\beta)^2}{2A_{2}}\right\}A_{2}^{-\frac{1}{2}}}.$\\

\end{enumerate}

\noindent Our goal is to estimate $\theta_{i}$, i.e., small area mean for the $i^{th}$ area, $i=1,\dots,m$. We implement Gibbs sampling using the conditional distributions (I)$-$(VI) in order to find posterior means and standard deviations of $\theta_{i}$'s. Conditional distribution (IV) and (V) may not have always admit a closed form expression.
%For,  $\sum\limits_{i=1}^{m}\delta_{i}=0$, $f(A_{1})\propto A_1^{-\alpha_{1}}I(A_{1}<A_{2})$. In that case  we may consider a transformation $U=\left( A_{1}/A_{2}\right)^{(1-\alpha_{1})}$ and based on that we generate $U$ from Uniform$(0,1)$ to obtain $A_{1}=A_{2}U^{\frac{1}{(1-\alpha_{1})}}$. [\textcolor[rgb]{1,0.4,0}{ I did not remove the following sentence}] \textcolor[rgb]{1,0,0}{[WHAT DO YOU MEAN BY THESE? When,  $\sum\limits_{i=1}^{m}\delta_{i}=1$ and $\alpha_{1}=0$ or $\sum\limits_{i=1}^{m}\delta_{i}=1/ 2$ and $\alpha_{1}=0$, acceptance-rejection with suitable chosen candidate density can be implemented.]} If  $\sum\limits_{i=1}^{m}\delta_{i}=1$, acceptance-rejection with suitably chosen candidate density can be implemented for all choices of $\alpha_{1}$. For $\sum\limits_{i=1}^{m}\delta_{i} \geq 2$ and $\alpha_{1}>0$ , given the other parameters, $A_{1}$ can be  generated from a truncated inverse-gamma distribution. Similarly, when $\sum\limits_{i=1}^{m}(1-\delta_{i})=0$, conditional distribution of $A_{2}$ is a Pareto distribution. Otherwise, given the other parameters, $A_{2}$ can be drawn from a truncated inverse-gamma distribution.

\subsection*{Proof of Theorem~\ref{thm:ch21}}\label{sec:ch2proof}

%Proof of Theorem~\ref{thm:ch21}:
Note that under the proposed mixture model, the likelihood function of the model parameter $\beta$, $A_{1}$, $A_{2}$ and $p$ based on the marginal distribution of $y_{1},\dots,y_{m}$ is given by
%\begin{align}
%\label{eqn:a1}
%L(\beta,A_{1},A_{2},p) & = C\times\prod_{i=1}^{m}\bigg[ \dfrac{p}{(A_{1}+D_{i})^{\frac{1}{2}}}\exp\left\{-\dfrac{(y_{i}-x^{T}_{i}\beta)^2}{2(A_{1}+D_{i})} \right\} \nonumber \\
%                       & ~~~~~~+  \dfrac{(1-p)}{(A_{2}+D_{i})^{\frac{1}{2}}}\exp\left\{-\dfrac{(y_{i}-x^{T}_{i}\beta)^2}{2(A_{2}+D_{i})} \right\}  \bigg],
%\end{align}
%\begin{align}
%\label{eqn:a1}
%L(\beta,A_{1},A_{2},p)  &= C\times\prod_{i=1}^{m}\bigg[ \dfrac{p}{(A_{1}+D_{i})^{\frac{1}{2}}}\exp\left\{-\dfrac{(y_{i}-x^{T}_{i}\beta)^2}{2(A_{1}+D_{i})} \right\} +  \dfrac{(1-p)}{(A_{2}+D_{i})^{\frac{1}{2}}}\exp\left\{-\dfrac{(y_{i}-x^{T}_{i}\beta)^2}{2(A_{2}+D_{i})} \right\}  \bigg],
%\end{align}
\begin{align}
\label{eqn:a1}
L(\beta,A_{1},A_{2},p)  &= C\times\prod_{i=1}^{m}\bigg[ \dfrac{p}{(A_{1}+D_{i})^{\frac{1}{2}}}e^{-\dfrac{(y_{i}-x^{T}_{i}\beta)^2}{2(A_{1}+D_{i})} } +  \dfrac{(1-p)}{(A_{2}+D_{i})^{\frac{1}{2}}}e^{-\dfrac{(y_{i}-x^{T}_{i}\beta)^2}{2(A_{2}+D_{i})} }  \bigg],
\end{align}
\noindent where $C$ is a generic positive constant not depending on the model parameters. Suppose for $0<a<b<\infty$ we have $a \le D_{i} \le b$, $i=1,\dots,m$. Since $(A_1+b) \ge (A_{1}+D_{i})\ge (a/b)(A_1+b)$,  $(A_2+b) \ge (A_{2}+D_{i})\ge (a/b)(A_2+b)$,
% $(A_{1}+D_{i})^{-1/2}$ is decreasing in $D_{i}$ and $\exp\left\{-\dfrac{(y_{i}-x^{T}_{i}\beta)^2}{2(A_{1}+D_{i})} \right\}$ is increasing in $D_{i}$,
from (\ref{eqn:a1})
%\begin{align}
%\label{eqn:a2}
%L(\beta,A_{1},A_{2},p) & \le C\times \prod_{i=1}^{m}\bigg[ \dfrac{p}{(A_{1}+a)^{\frac{1}{2}}}\exp\left\{-\dfrac{(y_{i}-x^{T}_{i}\beta)^2}{2(A_{1}+b)} \right\} \nonumber \\
%                       & ~~~~~~+  \dfrac{(1-p)}{(A_{2}+a)^{\frac{1}{2}}}\exp\left\{-\dfrac{(y_{i}-x^{T}_{i}\beta)^2}{2(A_{2}+b)} \right\}  \bigg].
%\end{align}
%\begin{align}
%\label{eqn:a2}
%L(\beta,A_{1},A_{2},p) & \le C\times \prod_{i=1}^{m}\bigg[ \dfrac{p}{(A_{1}+a)^{\frac{1}{2}}}\exp\left\{-\dfrac{(y_{i}-x^{T}_{i}\beta)^2}{2(A_{1}+b)} \right\} +  \dfrac{(1-p)}{(A_{2}+a)^{\frac{1}{2}}}\exp\left\{-\dfrac{(y_{i}-x^{T}_{i}\beta)^2}{2(A_{2}+b)} \right\}  \bigg].
%\end{align}
\begin{align}
\label{eqn:a2}
L(\beta,A_{1},A_{2},p) & \le C\times \prod_{i=1}^{m}\bigg[ \dfrac{p}{(A_{1}+b)^{\frac{1}{2}}}e^{-\dfrac{(y_{i}-x^{T}_{i}\beta)^2}{2(A_{1}+b)} } +  \dfrac{(1-p)}{(A_{2}+b)^{\frac{1}{2}}}e^{-\dfrac{(y_{i}-x^{T}_{i}\beta)^2}{2(A_{2}+b)} }  \bigg].
\end{align}
For $k=0,1,\cdots, m$, let $P_k = \{S_1^{(k)},S_2^{(k)}\}$ be an arbitrary partition of $\{1,2,\cdots,m\}$, where $S_1^{(k)}$ has $k$ elements and $S_2^{(k)}$ has $m-k= l (\mbox{say})$ elements. Let ${\cal P}_k$ denote all $m \choose k$ collections of $\{S_1^{(k)},S_2^{(k)}\}$.
%With this notation,
Then, expanding the product of the right hand side of (\ref{eqn:a2}), we get
\begin{align}
\label{eqn:a3}
L(\beta,A_{1},A_{2},p) & \le C\times \sum_{k=0}^{m}\sum_{P_k \in {\cal P}_k} \frac{p^k(1-p)^{m-k}}{(A_1+b)^{\frac k2} (A_2+b)^{\frac {m-k}2}} e^{-\sum_{i\in S_1^{(k)}} \dfrac{(y_{i}-x^{T}_{i}\beta)^2}{2(A_{1}+b)}  -\sum_{i\in S_2^{(k)}} \dfrac{(y_{i}-x^{T}_{i}\beta)^2}{2(A_{2}+b)}  }.
\end{align}
To establish propriety of the posterior density, we show integrability of each of the $2^m$ summands on the right hand side of (\ref{eqn:a3}) with respect to %the prior
$\pi(\beta,A_1,A_2,p)$ given in (\ref{eqn:pr1}).

\bigskip \noindent We first consider the case $k=0$. Here ${\cal P}_0$ has one element and $S^{(0)}$ is a null set. Let $Q(y)= y^T[I -X(X^TX)^{-1}X^T]y$.
%\bigskip \noindent
In this case, the integral $I^{(0)}$ of the term is
\begin{eqnarray}
I^{(0)} &=& C\int_0^\infty \int_{R^r} \int_0^{A_2}\int_0^1 (1-p)^mdp A_1^{-\alpha_1} dA_1 A_2^{-\alpha_2}(A_2+b)^{-\frac m2} e^{-\sum_{i=1}^m  \dfrac{(y_{i}-x^{T}_{i}\beta)^2}{2(A_{2}+b)}}d\beta dA_2 \nonumber
\end{eqnarray}

\begin{eqnarray}
&=& C \int_0^\infty \int_{R^r}   A_2^{1-\alpha_1-\alpha_2} (A_2+b)^{-\frac m2} e^{-\sum_{i=1}^m  \dfrac{(y_{i}-x^{T}_{i}\beta)^2}{2(A_{2}+b)}}d\beta dA_2 ~~~~~~(\mbox{since } \alpha_1 < 1)\nonumber\\
 &=& C \int_0^\infty A_2^{1-\alpha_1-\alpha_2} (A_2+b)^{-\frac {m-r}2} e^{-\frac 12 \frac{Q(y)}{A_2+b}} dA_2 \le C \int_0^\infty A_2^{1-\alpha_1-\alpha_2} (A_2+b)^{-\frac {m-r}2} dA_2 \nonumber\\
&<& \infty ,~~~~~\mbox{     if and only if } 2-\alpha_1-\alpha_2 > 0 \mbox{ and } 1-\alpha_1-\alpha_2 -\frac{m-r}2 < -1,
\label{t0}
\end{eqnarray}
which are equivalent to the conditions outlined in Theorem \ref{thm:ch21}.

\bigskip \noindent For the case $k=m$, again there is one term in ${\cal P}_m$ and the resulting integral, proceeding as in $I^{(0)}$, is bounded above by
\begin{eqnarray}
&& C\int_0^\infty A_1^{-\alpha_1} (A_1+b)^{-\frac{m-r}2}\int_{A_1}^\infty A_2^{-\alpha_2} dA_2  dA_1\nonumber\\
&=& C \int_0^\infty A_1^{1-\alpha_1-\alpha_2} (A_1+b)^{-\frac{m-r}2} dA_1 ~~(\mbox{since } \alpha_2>1) \nonumber\\
&<& \infty, ~~\mbox{under the conditions of the theorem}.
\label{tm}
\end{eqnarray}

\bigskip \noindent Now consider a case where $1\le k \le m-1$. Let $S_1^{(k)}$ be  a set of indices $\left\{i_{1},\dots,i_k \right\}$ and let $S_{2}^{(k)}=\{j_{1},\dots,j_{l}\} = \{1,2,\cdots,m\} \setminus S_1^{(k)}.$ Let us define, $M_{1}=\left(x_{i_{1}},\dots,x_{i_{k}} \right)^{T}$ and $M_{2}=\left(x_{j_{1}},\dots,x_{j_{l}} \right)^{T}$ . Suppose $g=\text{rank}(M_{1})$.  If  $g>0$, suppose $B \equiv \left\{\alpha_{1},\dots,\alpha_{g} \right\}\subset\left\{ i_{1},\dots,i_{k} \right\}$, so that $\left\{x_{\alpha_{1}},\dots,x_{\alpha_{g}} \right\}$ is linearly independent. If $g=0$, the set $B$ is always empty. Suppose $\left\{\gamma_{1},\dots,\gamma_{r-g} \right\}\subset \left\{ j_{1},\dots,j_{l} \right\}$ such that $\left\{x_{\alpha_{1}},\dots,x_{\alpha_{g}},x_{\gamma_{1}},\dots,x_{\gamma_{r-g}} \right\}$ is linearly independent. Let us define the $r \times r$ matrix $F = \left( x_{\alpha_{1}},\dots,x_{\alpha_{g}},x_{\gamma_{1}},\dots,x_{\gamma_{r-g}}\right)^{T}$, which is non-singular. Consider the non-singular linear transformation of $\beta$ by $\phi = F\beta$. With these developments, the integral of the term identified by $\{S_1^{(k)},S_2^{(k)}\}$ in the right hand side of (\ref{eqn:a3}) with respect to the prior $\pi(\beta,A_1,A_2,p)$ is bounded above by
\begin{eqnarray}
&& C\int_0^\infty \int_{A_1}^\infty \int_{R^r} \dfrac{A_{1}^{-\alpha_{1}}A_{2}^{-\alpha_{2}}}{{(A_{1}+b)^{\frac{k}{2}}(A_{2}+b)^{\frac{l}{2}}}} e^{-\sum_{i\in S_1^{(k)}} \dfrac{(y_{i}-x^{T}_{i}\beta)^2}{2(A_{1}+b)}  -\sum_{i\in S_2^{(k)}} \dfrac{(y_{i}-x^{T}_{i}\beta)^2}{2(A_{2}+b)}  }d\beta dA_2 dA_1\nonumber\\
&\le & C \int_0^\infty \int_{A_1}^\infty \int_{R^r} \dfrac{A_{1}^{-\alpha_{1}}A_{2}^{-\alpha_{2}}}{{(A_{1}+b)^{\frac{k}{2}}(A_{2}+b)^{\frac{l}{2}}}}
e^{-\sum_{u=1}^g \dfrac{(y_{\alpha_u}-x^{T}_{\alpha_u}\beta)^2}{2(A_{1}+b)}  -\sum_{t=1}^{r-g} \dfrac{(y_{\gamma_t} -x^{T}_{\gamma_t}\beta)^2} {2(A_{2}+b)}   } d\beta dA_2 dA_1\nonumber\\
&=&  C \int_0^\infty \int_{A_1}^\infty \int_{R^r} \dfrac{A_{1}^{-\alpha_{1}}A_{2}^{-\alpha_{2}}}{{(A_{1}+b)^{\frac{k}{2}}(A_{2}+b)^{\frac{l}{2}}}}
e^{-\sum_{u=1}^g \dfrac{(y_{\alpha_u}-\phi_u)^2}{2(A_{1}+b)}  -\sum_{t=1}^{r-g} \dfrac{(y_{\gamma_t} -\phi_{g+t})^2} {2(A_{2}+b)}   } d\phi dA_2 dA_1\nonumber
\end{eqnarray}

\begin{eqnarray}
&=& C  \int_0^\infty \int_{A_1}^\infty \dfrac{A_{1}^{-\alpha_{1}}A_{2}^{-\alpha_{2}}}{{(A_{1}+b)^{\frac{k-g}{2}}(A_{2}+b)^{\frac{l-r+g}{2}}}} dA_1 dA_2\nonumber\\
&\le& C  \int_0^\infty \int_{A_1}^\infty  \dfrac{A_{1}^{-\alpha_{1}}A_{2}^{-\alpha_{2}}}{{(A_{1}+b)^{\frac{k-g}{2}}(A_{1}+b)^{\frac{l-r+g}{2}}}} dA_2 dA_1 \nonumber\\
&=& C \int_0^\infty \dfrac{A_{1}^{1-\alpha_{1}-\alpha_{2}}}{{(A_{1}+b)^{\frac{k-g}{2}}(A_{1}+b)^{\frac{l-r+g}{2}}}} dA_1 \nonumber\\
&=& C \int_0^\infty \dfrac{A_{1}^{1-\alpha_{1}-\alpha_{2}}}{{(A_{1}+b)^{\frac{m-r}{2}}}} dA_1 <\infty,
\label{tk}
\end{eqnarray}
 by the conditions of the theorem.
Since the integrability conditions do not depend $k$ or on the indices $\left\{ i_{1},\dots,i_{k} \right\}$ and
$\left\{ j_{1},\dots,j_{l} \right\}$ and on the values $k$ and $l$, the conditions $2-\alpha_{1}-\alpha_{2}>0$ and $m>r+2(2-\alpha_{1}-\alpha_{2})$ will be sufficient to ensure the propriety of the posterior. $\square$

\end{document}